\documentclass[amsmath,amssymb,nofootinbib,superscriptaddress,showpacs,twocolumn]{revtex4}
\usepackage{graphicx}
\usepackage{dcolumn}
\usepackage{amsmath}
\usepackage[utf8]{inputenc}
\usepackage{amssymb}
\usepackage{natbib}
\usepackage{booktabs}
\usepackage{multirow}
\usepackage{tabularray}
\usepackage{gensymb}
\usepackage{lipsum}
\usepackage{blindtext}
\usepackage{mathtools}
\usepackage{booktabs}
\usepackage{tensor}
\usepackage[colorlinks=true, citecolor=blue, urlcolor = blue, linkcolor= red, bookmarks=true]{hyperref}
\usepackage{float}
\usepackage{amsfonts}
\usepackage{pgfplots}
\usepackage{mathrsfs}
\usepackage{afterpage}
\usepackage{xcolor}
\usepackage[dvipsnames]{xcolor}
\pgfplotsset{compat=1.10}
\begin{document}
\title[]{Probing Kalb–Ramond gravity with charged rotating black holes: constraints from EHT observations}
\author{Towheed Ahmad Nengroo} \email{tawheednengroo@gmail.com}
\affiliation{Centre for Theoretical Physics, 
	Jamia Millia Islamia, New Delhi 110025, India}
\author{Shafqat Ul Islam} \email{Shafphy@gmail.com}
\affiliation{Institute for Theoretical Physics and Cosmology, Zhejiang University of Technology, Hangzhou 310023, China}
\affiliation{Astrophysics and Cosmology Research Unit, 
	School of Mathematics, Statistics and Computer Science, 
	University of KwaZulu-Natal, Private Bag 54001, Durban 4000, South Africa}
\author{Sushant~G.~Ghosh }\email{sghosh2@jmi.ac.in}
\affiliation{Centre for Theoretical Physics, 
	Jamia Millia Islamia, New Delhi 110025, India}
\affiliation{Astrophysics and Cosmology Research Unit, 
	School of Mathematics, Statistics and Computer Science, 
	University of KwaZulu-Natal, Private Bag 54001, Durban 4000, South Africa}

\begin{abstract}
The Event Horizon Telescope (EHT) has guided strong-field gravitational physics by providing the first direct images of the supermassive black holes M87* and Sagittarius A*. The EHT observations offer unprecedented opportunities to test modified gravity theories against general relativity (GR). Motivated by this, we investigate charged rotating black holes in KR gravity, a framework motivated by string theory that incorporates spontaneous Lorentz symmetry breaking. The spacetime geometry is characterized by a Lorentz--violating parameter $\ell$ and electric charge $Q$, which modify the Kerr--Newman metric through a radial-dependent mass function. We compute black hole shadows and derive constraints on $\ell$ and $Q$ using EHT observations of M87* and Sgr A*. For angular shadow diameter $\theta_{\rm sh}$ of M87* at inclination $\theta_o=17^\circ$ and fixed $Q=0.2$, the EHT-allowed range $\theta_{\rm sh}\in(35.1,\,40.5)\,\mu\mathrm{as}$ constrains the Lorentz--violating parameter to approximately $-0.019\lesssim\ell\lesssim0.075$ and $-0.076\lesssim\ell\lesssim0.029$ across the admissible spin interval. For angular shadow diameter $\theta_{\rm sh}$ of Sgr A* at inclination $\theta_o=50^\circ$ and fixed $Q=0.2$, the corresponding EHT-allowed range $\theta_{\rm sh}\in(41.7,\,55.7)\,\mu\mathrm{as}$ permits approximately $-0.075\lesssim\ell\lesssim0.110$ and $-0.124\lesssim\ell\lesssim0.076$ across the admissible spin interval. Our analysis reveals that the Lorentz-violating parameter suppresses the shadow radius by a factor $\sqrt{1-\ell}$, while charge introduces additional distortions. Using the angular shadow diameter measured by EHT, we obtain an upper bound $\ell \lesssim 0.19$ from Sgr A* data with the stellar dynamics mass prior. These results demonstrate the power of EHT observations to constrain Planck-scale Lorentz violation and provide insights into the fundamental nature of spacetime.
\end{abstract}
\keywords{Galaxy: center–
	gravitation – black hole physics -black hole shadow-  gravitational lensing: strong}
\maketitle
\section{Introduction}\label{Sec-1}
General relativity (GR) has been incredibly successful in describing gravitational phenomena from weak-field tests such as the perihelion precession of Mercury to strong-field observations, including the recent imaging of black hole shadows by the Event Horizon Telescope (EHT) \cite{EventHorizonTelescope:2019ggy, EventHorizonTelescope:2019pgp, EventHorizonTelescope:2022wkp, EventHorizonTelescope:2022exc}. Despite the successes of GR, it remains inconsistent with quantum gravity, and increasing observational precision motivates the exploration of modified theories of gravity that may demonstrate deviations from GR \cite{Clifton:2011jh, Will:2014kxa}. Among the several extensions proposed, those incorporating Lorentz symmetry breaking have attracted particular attention, as violations of Lorentz invariance emerge naturally in various quantum gravity approaches, including string theory, loop quantum gravity, Hořava–Lifshitz gravity, and noncommutative field theories \cite{Kostelecky:2008ts, Mattingly:2005re, Horava:2009uw}.

The Kalb–Ramond (KR) field is an antisymmetric tensor field that naturally appears in bosonic string theory \cite{Kalb:1974yc, Scherk:1974ca}, which is a useful framework for studying possible violations of Lorentz symmetry. If the KR field acquires a nonzero vacuum expectation value via spontaneous symmetry breaking, it introduces a preferred direction in spacetime. Even with this preferred direction, the theory still remains invariant under observer Lorentz transformations \cite{Kostelecky:2003cr}. When the KR field is coupled to gravity in a non-minimal way, it alters the usual dynamics of gravity. In particular, the antisymmetric tensor field can contribute to spacetime torsion and may lead to observable physical effects \cite{Altschul:2009ae, Duan:2023gng, Kumar:2020hgm}.

Lessa and Maluf \cite{Lessa:2019bgi} first obtained the exact Schwarzschild black hole in KR gravity. Subsequent investigations exploring particle motion, gravitational lensing, and energy extraction mechanisms \cite{Jumaniyozov:2025lox, Yao:2026cvs}. The Reissner--Nordstr\"{o}m black hole in the theory exhibited that Lorentz violation significantly modifies black hole thermodynamics, phase structure, and stability criteria, while preserving the standard first law and Smarr formula. Significant attention has been given to black holes in KR gravity, analysing their thermodynamic properties and quasi-periodic oscillations \cite{Jumaniyozov:2025dyy}, as well as the dynamics of charged and spinning particles in these backgrounds \cite{Jumaniyozov:2025lox, Jumaniyozov:2025nnj, Zahid:2024ohn}. Furthermore, the influence perfect fluid dark matter, plasma media, and nonlinear electrodynamics—has been rigorously analyzed to determine their impact on gravitational lensing \cite{Rahmatov:2025gpk, Rahmatov:2026clv, Rahmatov:2026mdo, Saydullayev:2025oop} and thermodynamic stability \cite{Rahmatov:2025ari, Sekhmani:2025jbl}.

\cite{Kanzi:2019gtu,Maluf:2020kgf}. While the rotating  Kerr-like  solutions constructed via the Newman–Janis algorithm have enabled studies of ergoregions, frame-dragging effects, and the impact of the Lorentz-violating parameter on horizon structure \cite{Kumar:2020cve, Kumar:2020hgm, Kumar:2020ltt}. The recent observational breakthroughs by the EHT – providing the first direct images of the supermassive black holes M87* and Sgr A* – have inaugurated a new era of strong-field gravitational physics \cite{EventHorizonTelescope:2019dse, EventHorizonTelescope:2022xnr}. Horizon-scale observations provide an important way to test modified gravity models by comparing the predicted shadow size and shape with observational measurements \cite{Kumar:2018ple}. The black hole shadow, which represents the photon capture region on the observer’s sky, carries useful information about the spacetime geometry and can therefore be used to study possible deviations from the Kerr solution of general relativity \cite{Falcke:1999pj, Bardeen:1973tla, Psaltis:2018xkc, Kumar:2018ple, Kumar:2020hgm}. It appears as a dark region because some light rays are captured by the black hole and do not reach a distant observer. This feature was first studied in a systematic way by Falcke {\it et al.} \cite{Falcke:1999pj}, who predicted that the supermassive black hole at the center of our Galaxy would cast an observable shadow. Later, Huang et al. \cite{Huang:2007us} carried out a detailed visibility analysis for Sagittarius A*, demonstrating how interferometric observations could resolve the black hole shadow. These foundational studies laid the groundwork for the remarkable achievements of the EHT collaboration, which has since produced the first direct images of the supermassive black holes M87* and Sgr A* \cite{EventHorizonTelescope:2019dse, EventHorizonTelescope:2019ggy, EventHorizonTelescope:2022wkp, EventHorizonTelescope:2024dhe, EventHorizonTelescope:2025dua}, providing a new opportunity for testing gravitational theories in the strong-field regime.

The shadow of a black hole contains crucial information about the underlying spacetime geometry and provides a strong method for finding deviations from the Kerr metric of general relativity. Amarilla, Eiroa, and their collaborators were the first to study the black hole shadows in modified gravity theories, including rotating black holes in extended Chern-Simons modified gravity \cite{Amarilla:2010zq}, braneworld scenarios \cite{Amarilla:2011fx, Eiroa:2017uuq}, and Kaluza-Klein rotating dilaton black holes \cite{Amarilla:2013sj}. Atamurotov {\it et al.} extended this to Ho\v{r}ava-Lifshitz gravity \cite{Atamurotov:2013dpa}, non-Kerr black holes \cite{Atamurotov:2013sca}, Kerr-Taub-NUT spacetimes \cite{Abdujabbarov:2012bn}, and 5D Myers-Perry black holes \cite{Papnoi:2014aaa}. Abdujabbarov {\it et al.} \cite{Abdujabbarov:2015xqa} introduced a coordinate-independent formalism for describing shadow boundaries. Tsukamoto, Li, and Bambi \cite{Tsukamoto:2014tja, Tsukamoto:2017fxq, Li:2013jra} did systematic studies into relationship between shadow observables and black hole parameters, demonstrating how spin and deformation parameters can be constrained from shadow measurements.

The characterization of shadow observables has been refined through numerous studies. Wei, Liu, and their collaborators \cite{Wei:2013kza, Wei:2015dua, Wei:2019pjf, Wei:2020ght} explored the connection between shadow radius and curvature properties of spacetime, as well as shadows in noncommutative geometry inspired black holes and Gauss-Bonnet gravity. Perlick {\it et al.} \cite{Perlick:2015vta, Perlick:2018iye, Bisnovatyi-Kogan:2017kii, Bisnovatyi-Kogan:2018vxl, Tsupko:2017rdo, Tsupko:2018apb, Tsupko:2019pzg} made seminal contributions to understanding plasma effects on black hole shadows, showing that the presence of a surrounding plasma can significantly alter shadow size and shape. Cunha {\it et al.} \cite{Cunha:2018gql, Cunha:2019hzj} investigated the connection between event horizon and shadow geometry. While Chael {\it et al.} \cite{Chael:2021rjo} developed methods for seeing the inner shadow, Dokuchaev and Nazarova \cite{Dokuchaev:2018kzk} investigated event horizon imaging within black hole shadows.

To bridge the gap between theoretical models and EHT data, several investigation have utilized EHT observations of M87* and Sgr A* to place constraints on black hole shadows and quasinormal modes within Gauss-Bonnet, Horndeski, and ModMax theories \cite{Zahid:2024hyy, Zahid:2022eeq, Zahid:2023csk, Raza:2023vkn, Rayimbaev:2022znx, Zahid:2024hwi, Zubair:2023cep, Khan:2023qsl, Sekhmani:2024fjn, Sekhmani:2026iyg, Rahmatov:2025dlm}. Collectively, these works demonstrate that the coupling of modified gravity with non-trivial matter sources provides a fertile ground for testing the limits of the current cosmological paradigm. Psaltis, \"Ozel, and their collaborators \cite{Psaltis:2014mca, Psaltis:2018xkc, EventHorizonTelescope:2020qrl} developed null hypothesis tests for general relativity using EHT observations. Kumar and Ghosh \cite{Kumar:2018ple, Kumar:2019ohr, Kumar:2020ltt, Kumar:2020owy, Ghosh:2020spb} developed parameter estimation techniques for black hole shadows and studied rotating black holes in asymptotically safe gravity. Vagnozzi {\it et al.} \cite{Vagnozzi:2022moj, Vagnozzi:2022tba} review paper furnished horizon-scale tests of gravity theories using the Sgr A* image. Ghosh and Afrin \cite{Ghosh:2022kit, Afrin:2021wlj, Afrin:2023uzo} derived upper limits on black hole charge and tested Horndeski gravity using EHT results. Kumar {\it et al.} \cite{KumarWalia:2022aop, Abdujabbarov:2016hnw} tested rotating regular metrics with EHT observations. Ayzenberg and Yunes \cite{Ayzenberg:2018jip, Ayzenberg:2022twz} tested GR with black hole shadows in quadratic gravity, while Ayzenberg {\it et al.} \cite{Ayzenberg:2023hfw} outlined fundamental physics opportunities with future VLBI arrays. Afrin {\it et al.} \cite{Afrin:2022ztr} performed tests of LQG using EHT results for Sgr A*, while Islam et al. \cite{Islam:2022wck, Kumar:2023jgh} investigated LQG motivated rotating black holes and their strong gravitational lensing signatures. Ali {\it et al.} \cite{Ali:2024ssf} studied shadows and parameter estimation of rotating quantum-corrected black holes. Sekhmani et al. \cite{Sekhmani:2025bsi} extended this analysis to charged rotating nonsingular black holes in loop quantum gravity. Liu et al. \cite{Liu:2020ola} studied shadow and quasinormal modes of rotating loop quantum black holes.

Given the theoretical motivation for Lorentz-violating gravity from string theory and the unprecedented observational capabilities of the EHT, a natural question arises: Can current horizon-scale images of M87* and Sgr A* constrain the Lorentz-violating parameter $\ell$ in KR gravity? And more broadly, do charged rotating black holes in KR gravity produce shadow features that are observationally distinguishable from their Kerr counterparts?

In this paper, we address these questions by systematically investigating the shadow properties of charged rotating black holes in KR gravity. The spacetime geometry under consideration is characterized by three parameters: the spin $a$, the electric charge $Q$, and the Lorentz-violating parameter $\ell$. Using the modified Newman-Janis algorithm, we construct the rotating solution and analyze the resulting null geodesic structure, the photon region, and the shadow in celestial coordinates. We compute key shadow observables, including the shadow area $A$ and oblateness $D$, following the robust parameter estimation framework developed by Ghosh and Kumar \cite{Kumar:2018ple, Kumar:2020owy}. Using these observables, we derive constraints on $\ell$ and $a$ by comparing with the angular shadow diameters reported by the EHT for M87* and Sgr A* \cite{EventHorizonTelescope:2019dse, EventHorizonTelescope:2022xnr}. 

This paper is organized as follows:  In Section~\ref{Sec-2}, we present the charged rotating black hole solution in KR gravity, describing the effective action, the role of Lorentz symmetry breaking, and the construction of the rotating metric via the modified Newman-Janis algorithm. Section~\ref{Sec-3}  is devoted to the black hole shadow, deriving the null geodesic equations, the photon region, and the celestial coordinates that determine the shadow's apparent shape. Section~\ref{Sec-4} describes the parameter estimation methodology, introducing shadow observables such as area and oblateness, and discusses the thermal energy emission spectrum. Section~\ref{Sec-5} is devoted to analyzing EHT observations of M87* and Sgr A* to derive observational constraints on the Lorentz-violating parameter $\ell$ and spin $a$. Finally, in Section~\ref{Sec-6}, we conclude with a summary of the results, emphasizing the consistency of KR black holes with current EHT data and the prospects for future tests with ngEHT.

\begin{figure*}
\centering
\begin{tabular}{c c}
\includegraphics[scale=0.7]{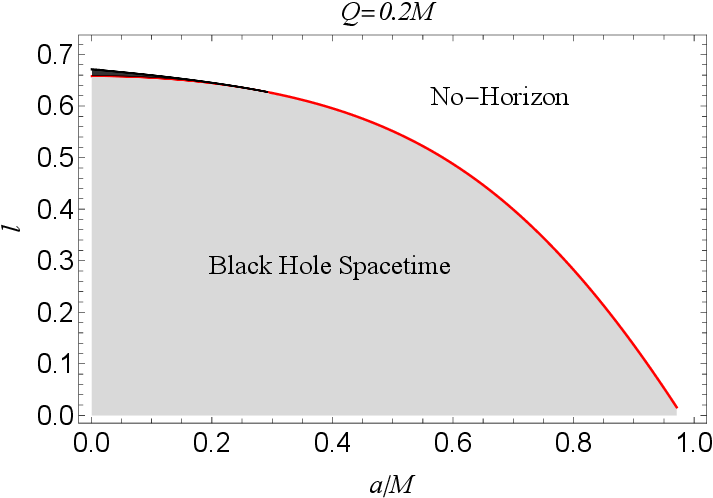} 
\hspace{1cm} &
\includegraphics[scale=0.7]{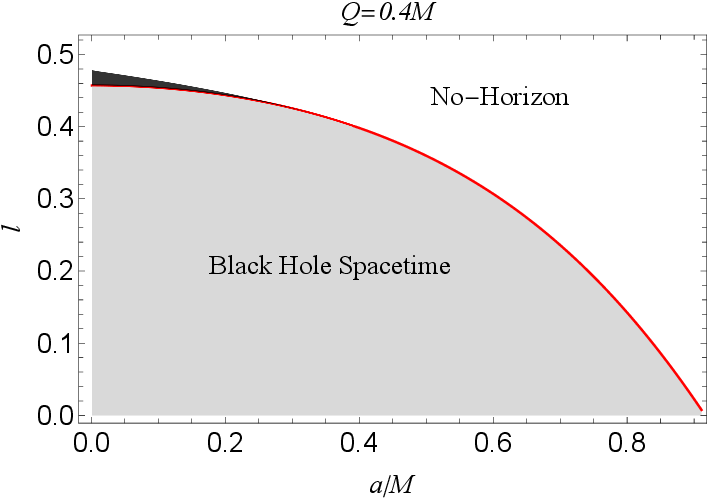}
\end{tabular}
 
   \caption{Parameter space in the $(a/M,\ell)$ plane for a rotating black hole in KR gravity at fixed charges $Q = 0.2M$ (left), and $Q = 0.4M$ (right). The red curve separates the boundary between black hole spacetimes and no-horizon configurations. Within the horizon regime, the black shaded region indicates the subset of parameter space that admits closed photon rings.}\label{parameter}
\end{figure*}

\section{Charged rotating BH in KR gravity}\label{Sec-2}
The charged and rotating black hole solutions in the presence of a KR field can be derived from the following effective action
 \begin{eqnarray}\label{eq1}
S_{\mathrm{KR}} &=& \frac{1}{2}\int d^{4}x\sqrt{-g}
\Big[
R-\frac{1}{6}H_{\mu\nu\sigma}H^{\mu\nu\sigma}
-V(B_{\mu\nu}B^{\mu\nu}\pm b^{2})
\nonumber\\
&&+\xi_{2}B_{\sigma\mu}B^{\nu\mu}R^{\sigma}{}_{\nu}
+\xi_{3}B_{\mu\nu}B^{\mu\nu}R
\Big],
\end{eqnarray}
 where $g$ is the determinant of the metric $g_{\mu\nu}$, $R_{\mu\nu}$ is the Ricci tensor, and $R$ is the Ricci scalar. The antisymmetric tensor field $B_{\mu\nu}$ represents the KR field originally arising in the low-energy limit of string theory~\cite{Kalb:1974yc,Green:1987sp,Polchinski:1998rq}. The field strength is defined as $H_{\mu\nu\sigma}=\partial_{[\mu}B_{\nu\sigma]}$, which is invariant under the gauge transformation $B_{\mu\nu}\rightarrow B_{\mu\nu}+\partial_{[\mu}\Lambda_{\nu]}$. We have introduced the potential $V(B_{\mu\nu}B^{\mu\nu}\pm b^{2})$ for a nonzero vacuum expectation value for the KR field, leading to spontaneous Lorentz symmetry breaking, analogous to bumblebee-type models~\cite{Kostelecky:1989jp,Kostelecky:2003fs,Bluhm:2004ep}. The parameters $\xi_{2}$ and $\xi_{3}$ represent the nonminimal couplings between the KR field and the spacetime curvature, which naturally arise in effective field theory descriptions of Lorentz-violating gravity~\cite{Bailey:2006fd,Bambi:2015kza,Jumaniyozov2025}.
In addition to the Einstein–Hilbert term, the action in Eq.~(\ref{eq1}) therefore incorporates both the kinetic and self-interaction contributions of the KR field, along with its coupling to curvature. A nonvanishing background value of $B_{\mu\nu}$ selects a preferred direction in spacetime, thereby breaking local Lorentz invariance while preserving general covariance.
For physical interpretation, it is often convenient to decompose the antisymmetric tensor $B_{\mu\nu}$ into pseudo-electric and pseudo-magnetic components with respect to a timelike four-vector $v^{\mu}$~\cite{Maluf:2020kgf}. In this decomposition, the resulting spacelike vectors are orthogonal to $v^{\mu}$ and can be interpreted analogously to electric and magnetic fields in classical electrodynamics, providing further insight into the dynamical role of the KR field in gravitational systems.

To construct electrically charged solutions in gravitational theories that include a KR tensor field, 
the matter sector is extended to contain the standard electromagnetic field $A_\mu$, with field strength 
$F_{\mu\nu}=2\nabla_{[\mu}A_{\nu]}$, together with a nonminimal interaction involving the KR field $B_{\mu\nu}$ 
and curvature. The electromagnetic field enters the action in the usual Maxwell form, minimally coupled to the 
metric, while the KR field strength $H_{\lambda\mu\nu}=\partial_{[\lambda}B_{\mu\nu]}$ is defined analogously 
to the three-form field strength in string theory--inspired models \cite{Maluf:2020kgf}.
The self-interaction potential 
$V(B_{\mu\nu}B^{\mu\nu}\pm b^2)$ depends only on the Lorentz-invariant contraction $B_{\mu\nu}B^{\mu\nu}$. 
A smooth potential of this form is introduced to drive spontaneous Lorentz symmetry breaking --- preserving 
\emph{observer} Lorentz invariance in the fundamental Lagrangian, while allowing the vacuum to choose a 
preferred tensor direction \cite{Lessa:2019bgi, Duan:2023gng, Junior:2024ety}. The potential is constructed so 
that it attains its minimum when
\begin{equation}
B_{\mu\nu}B^{\mu\nu}=\mp\,b^2,
\end{equation}
with the sign chosen such that $b^2>0$. At this minimum, the potential vanishes in the absence of a cosmological 
constant and the KR field acquires a nonzero vacuum expectation value (VEV)
\begin{equation}
\langle B_{\mu\nu}\rangle=b_{\mu\nu},
\end{equation}
which acts as a fixed background tensor that spontaneously breaks local Lorentz symmetry \cite{Lessa:2019bgi, Duan:2023gng}.
Because the KR field couples nonminimally to curvature --- for example through terms like 
$\varepsilon\,B^{\mu\lambda}B^\nu{}_\lambda R_{\mu\nu}$ in the action --- the vacuum configuration $b_{\mu\nu}$ 
influences the gravitational dynamics. This nonminimal coupling means that the KR condensate does not simply 
decouple; instead it backreacts on the spacetime geometry and induces spacetime anisotropies and breaking of 
local Lorentz invariance in particle dynamics, leading to modified black hole and charged solutions with 
distinctive observational signatures \cite{Duan:2023gng, Sucu:2025lqa}.

Following earlier studies, we assume that the only nonvanishing components of the vacuum configuration of the KR field are \cite{Duan:2023gng}
\begin{eqnarray}
b_{10}=-b_{01}=\tilde{E}(r),
\label{02}
\end{eqnarray}

With the vacuum configuration specified in Eq.~(\ref{02}), the KR field strength vanishes identically and therefore does not contribute through its kinetic term. To describe electrically charged configurations, we introduce an electrostatic vector potential of the form $A_{\mu}=-\Phi(r)\,\delta^{t}_{\mu}$, which is compatible with the static and spherically symmetric spacetime.

In the presence of a nontrivial KR background, the electromagnetic field alone is not sufficient to support charged black hole solutions. An additional interaction between the electromagnetic field and the KR field is therefore required. This interaction is incorporated directly to the matter Lagrangian. Accordingly, we take the matter Lagrangian to be
\begin{eqnarray}
L_M=-\frac{1}{2}F_{\mu\nu}F^{\mu\nu}
-\eta B^{\alpha\beta}B^{\gamma\rho}F_{\alpha\beta}F_{\gamma\rho},
\label{03}
\end{eqnarray}
where $\eta$ is a coupling parameter mediating Lorentz symmetry breaking in the gauge field through the background KR configuration. When the KR field acquires a nonzero vacuum expectation value, this interaction modifies the electromagnetic field equations, yielding electrically charged black hole solutions.

The modified Einstein equations are obtained by varying the total action in Eq.~(\ref{eq1}) with respect to the metric $g_{\mu\nu}$. 
This yields the field equations
\begin{eqnarray}
R_{\mu\nu}-\frac{1}{2}g_{\mu\nu}R
= T^{M}_{\mu\nu}+T^{\mathrm{KR}}_{\mu\nu},
\label{04}
\end{eqnarray}
where $T^{M}_{\mu\nu}$ denotes the energy--momentum tensor associated with the electromagnetic field, while 
$T^{\mathrm{KR}}_{\mu\nu}$ accounts for the contributions arising from the KR field and its nonminimal couplings 
to gravity \cite{Duan:2023gng}.
The energy--momentum tensor of the electromagnetic field is derived from the matter Lagrangian in Eq.~(\ref{03}) and takes the explicit form
\begin{eqnarray}
T^{M}_{\mu\nu}
&=& 2F_{\mu\alpha}F_{\nu}{}^{\alpha}
-\frac{1}{2}g_{\mu\nu}F_{\alpha\beta}F^{\alpha\beta} \nonumber\\
&& + \eta \Big(
8B^{\alpha\beta}B_{\nu\gamma}F_{\alpha\beta}F_{\gamma\mu}
- g_{\mu\nu}B^{\alpha\beta}B^{\gamma\delta}
F_{\alpha\beta}F_{\gamma\delta}
\Big).\quad \quad
\label{05}
\end{eqnarray}
The terms proportional to $\eta$ encode the interaction between the electromagnetic field and the KR background, 
effectively modifying the dynamics of the gauge field in the presence of the vacuum KR configuration \cite{Sucu:2025lqa}.
\begin{figure}[t]
\centering
\includegraphics[scale=0.65]{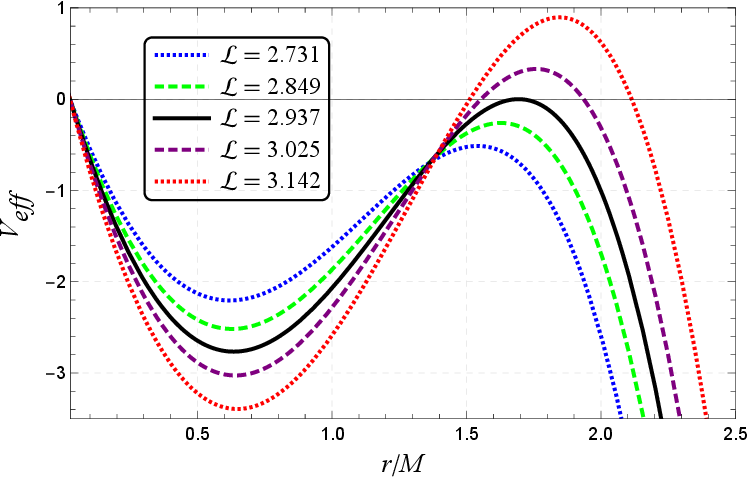}
\caption{Effective potential $V_{\mathrm{eff}}$ as a function of $r/M$ for equatorial photon motion $(\eta=0)$, shown for different values of the angular momentum $\mathcal{L}$ with fixed parameters  $a=0.7M$, $\ell=0.1$, and $Q=0.2M$. The curves correspond to $\mathcal{L} = 0.93\,\mathcal{L}_c$, $0.97\,\mathcal{L}_c$, $\mathcal{L}_c$, $1.03\,\mathcal{L}_c$, and $1.07\,\mathcal{L}_c$, where $\mathcal{L}_c=2.937$ denotes the critical angular momentum.}
\label{fig:Veff}
\end{figure}
\begin{figure*}[t]
    \begin{tabular}{c c}
    \includegraphics[scale=0.55]{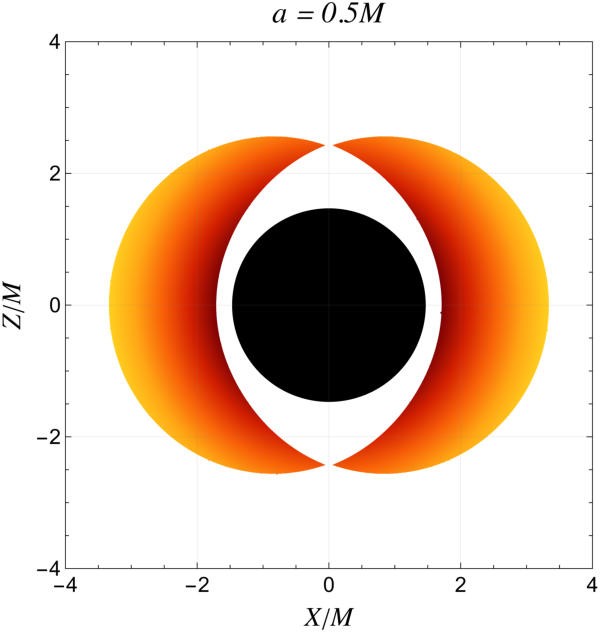}
     \includegraphics[scale=0.55]{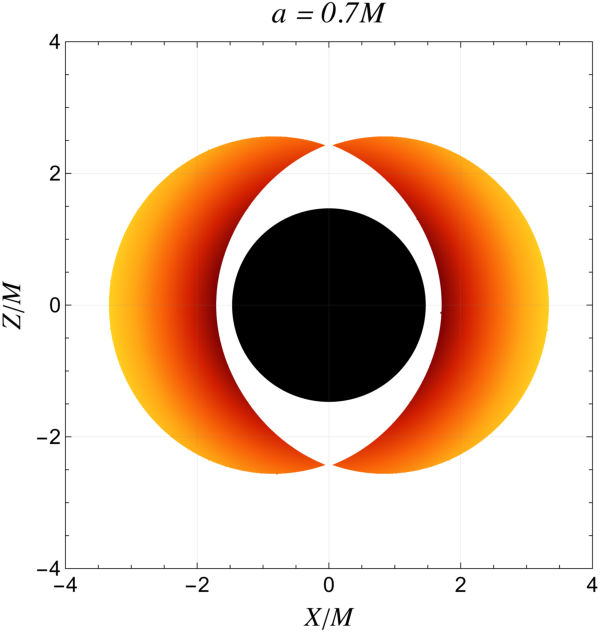}
      \includegraphics[scale=0.55]{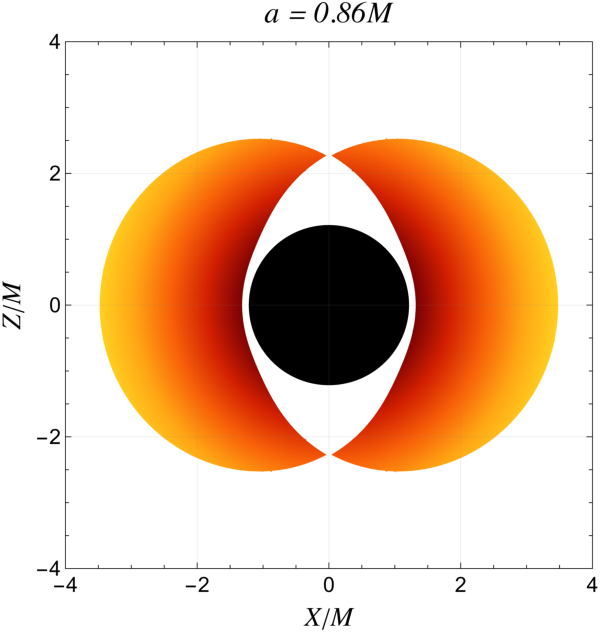}
   \end{tabular}
   \caption{Configuration of the spherical photon region (shaded red area), with the event horizon shown as the black disk. The plot represents a cross-section in the $(x,z)$ plane constructed from the Boyer-Lindquist coordinates via $x = r\sin \theta$ and $z = r\cos \theta$. The spin parameter increases from left to right, $a = 0.5M,~0.7M,~0.86M$, while $\ell = 0.1$ and $Q = 0.2M$ are held fixed.}\label{figPRP}
\end{figure*}
The effective energy--momentum tensor associated with the KR field is given by
\begin{eqnarray}
&&T^{\mathrm{KR}}_{\mu\nu}
=\frac{1}{2}H_{\mu\alpha\beta}H_{\nu}{}^{\alpha\beta}
-\frac{1}{12}g_{\mu\nu}H_{\alpha\beta\gamma}H^{\alpha\beta\gamma}
+2V'B_{\mu\alpha}B_{\nu}{}^{\alpha}
\nonumber\\ && - g_{\mu\nu}V +\xi_{2}\Big[
\frac{1}{2}g_{\mu\nu}B^{\alpha\beta}B^{\gamma\delta}R_{\alpha\gamma}
-B^{\alpha}{}_{\mu}B^{\beta}{}_{\nu}R_{\alpha\beta}
\nonumber\\&&-B^{\alpha\beta}B_{\mu}{}^{\gamma}R_{\alpha\gamma\nu\beta}
\Big]-\xi_{3}\Big[
B_{\mu\alpha}B_{\nu}{}^{\alpha}R
+\frac{1}{2}g_{\mu\nu}B^{\alpha\beta}B_{\alpha\beta}R 
\nonumber\\&&-\nabla_{\alpha}\nabla_{\mu}(B^{\alpha\beta}B_{\nu\beta})
-\nabla_{\alpha}\nabla_{\nu}(B^{\alpha\beta}B_{\mu\beta})
\Big],
\label{06}
\end{eqnarray}
The prime denotes differentiation with respect to the argument of the potential. It is worth noting that the total energy--momentum tensor, given by the sum $T^{M}_{\mu\nu}+T^{\mathrm{KR}}_{\mu\nu}$, is covariantly conserved as a consequence of the Bianchi identities. 

\begin{eqnarray}
\nabla^{\nu}\!\left(
F_{\nu\mu}
+2\eta B_{\nu\alpha}B_{\mu\beta}F^{\alpha\beta}
\right)=0 .
\label{08}
\end{eqnarray}
When the coupling constant $\eta$ is set to zero, Eq.~(\ref{08}) reduces to the conventional Maxwell equation in curved spacetime.  Finally, the charged non-rotating KR BH  derived using the field equations, has  the following form \cite{Jumaniyozov:2025lox}: 
\begin{eqnarray}
ds^{2}
=-f(r)dt^{2}
+g(r)dr^{2}
+r^{2}\left(d\theta^{2}+\sin^{2}\theta\,d\phi^{2}\right).
\label{09}
\end{eqnarray}
\begin{eqnarray}
f(r)=\frac{1}{g(r)}
=\frac{1}{1-\ell}-\frac{2M}{r}
+\frac{Q^{2}}{(1-\ell)^{2}r^{2}},
\label{24}
\end{eqnarray}

where, $M$ denotes the mass of the black hole and $Q$ represents its electric charge and $\ell$ is the Lorentz-violating parameter. In the limit $\ell\rightarrow 0$, the metric reduces to the Reissner--Nordström solution. Furthermore, when both $\ell=0$ and $Q=0$, the Schwarzschild spacetime is recovered. We acknowledge that $f(r) \to 1/(1-l)$ as $r \to \infty$ \cite{Horava:2009uw}. However, for physically realistic scenarios where the KR parameter is very small ($l \ll 1$), the metric effectively recovers the standard Minkowski limit, $f(r) \approx 1$. The presence of the $1/(1-l)$ factor is a foundational feature of spontaneous Lorentz symmetry breaking that we have now formally addressed.

To obtain the rotating counterpart of the static charged solution, we employ the modified Newman--Janis algorithm (NJA), 
which provides a systematic prescription for generating axially symmetric spacetimes from spherically symmetric seed metrics. 
This procedure has been widely used to construct rotating solutions in Kerr-like and modified gravity models 
\cite{Newman:1965tw, Ghosh:2014pba, Kumar:2019ohr, Kumar:2018ple, Kumar:2020ltt, Kumar:2025bim}. 
In particular, we obtain its rotating counterpart using the modified Newman--Janis algorithm (MNJA) 
based on Azreg-A\"inou’s non-complexification procedure \cite{Azreg-Ainou:2014aqa, Azreg-Ainou:2014pra}. 
In Boyer--Lindquist coordinates $(t, r, \theta, \phi)$, the charged rotating KR black hole metric reads
\begin{eqnarray}
ds^{2} &=& -\left(1-\frac{2M(r)r}{\Sigma}\right)dt^{2}
+\frac{\Sigma}{\Delta}\,dr^{2}
+\Sigma\,d\theta^{2}\nonumber\\
&&
-\frac{4aM(r)r\sin^{2}\theta}{\Sigma}\,dt\,d\phi
+\sin^{2}\theta\nonumber\\
&&\left(r^{2}+a^{2}
+\frac{2a^{2}M(r)r\sin^{2}\theta}{\Sigma}\right)d\phi^{2}
\label{28}
\end{eqnarray}
where
\begin{eqnarray}
\Sigma &=& r^{2}+a^{2}\cos^{2}\theta , \nonumber\\
\Delta &=& r^{2}+a^{2}-2M(r)r , \nonumber\\
M(r) &=& M-\frac{\ell\,r}{2(1-\ell)}
-\frac{Q^{2}}{2(1-\ell)^{2}r}.
\label{29}
\end{eqnarray}
Here, $a$ denotes the rotation parameter, while $\Sigma$ and $\Delta$ characterize the angular and radial 
structure of the rotating spacetime, respectively. The parameter $\ell$ represents the Lorentz-violating scale arising from the KR vacuum configuration and enters the rotating geometry 
through the mass function $M(r)$, which encodes both charge and Lorentz‑violating effects \citep{Duan:2023gng}.
The Newman--Janis algorithm (NJA) has long served as the primary technique for generating rotating black hole solutions from their static, spherically symmetric counterparts \cite{Newman:1965tw}. While the NJA success in deriving the Kerr and Kerr--Newman metrics within GR is often regarded as a mathematical ``miracle'' \cite{Newman:1965tw,Azreg-Ainou:2014pra}, its application to modified gravity (MOG) theories presents significant challenges. It has been rigorously demonstrated that applying the standard NJA complexification to arbitrary non-GR seeds frequently introduces physical pathologies, such as ``ghost'' sources or a failure to satisfy the underlying field equations of the theory \cite{Hansen:2013owa}. In the GR framework, the NJA preserves the source identity---e.g., transitioning from a vacuum Schwarzschild seed to a vacuum Kerr solution. However, in modified gravity, the standard NJA often yields rotating counterparts that necessitate additional stresses \cite{Cirilo-Lombardo:2021ouj}. 

To circumvent these inconsistencies and the ambiguities inherent in the complexification of radial coordinates, we employ the Modified Newman--Janis Algorithm (mNJA) based on Azreg-A\"inou's non-complexification procedure \cite{Azreg-Ainou:2014pra, Azreg-Ainou:2014aqa}. This refined approach has been successfully utilized to generate rotating solutions in diverse modified gravities \cite{Kumar:2017qws,Kumar:2020owy,Kumar:2020hgm,Ghosh:2014pba,Ghosh:2015ovj,Walia:2021emv,Ahmed:2025zdc}, including $f(R)$ gravity \cite{delaCruz-Dombriz:2009pzc,Ghosh:2014hea} and LQG-inspired models \cite{Brahma:2020eos,Kumar:2022vfg,Ali:2024ssf}. By allowing for the presence of imperfect fluid components and ensuring the separability of the Hamilton--Jacobi equations, the mNJA provides a consistent path for generating rotating Einstein--Gauss--Bonnet (EGB) black holes in Boyer--Lindquist-like coordinates \cite{Kumar:2020owy,Toshmatov:2017bpx}. Consequently, this procedure ensures that the resulting axially symmetric metric remains a mathematically viable and physically regular solution within the EGB framework.

The horizon structure of the charged rotating KR black hole is determined by the roots of the  radial function $\Delta(r)=0$, where $\Delta$ is defined in Eq.~(\ref{29}). This yields
\begin{equation}
r_{\pm} =   M(1 - \ell) \pm \sqrt{(1 - \ell)\left(M^2(1 - \ell) - a^2 - \frac{Q^2}{1 - \ell}\right)}
\end{equation}
with $r_+$ and $r_-$ denoting the event and Cauchy horizons, respectively \citep{Newman:1965tw}.  In the Kerr limit of the theory, obtained by setting the Lorentz‑violating parameter $\ell=0$ and the 
electromagnetic charge $Q=0$, the horizon radii reduce to their exact analytical expressions for the Kerr 
black hole \citep{Kerr:1963ud, Bardeen:1972fi}:
\begin{equation}
r_{\pm}^{K} = M \pm \sqrt{M^{2} - a^{2}}
\end{equation}

These expressions represent the 
classical limit in which frame dragging dominates the near‑horizon structure. 
In the full modified KR geometry, depending on the values of the parameters $(a,\ell,Q)$, the function  $\Delta(r)$ may admit two distinct real roots (non‑extremal black hole), a degenerate root (extremal  configuration), or no real roots at all (horizonless configuration). Such behavior, where additional  theory parameters shift and split horizon radii relative to the Kerr case, is a generic feature of black 
holes in Lorentz‑violating \footnote{Points to an example in Lorentz-violating gravity where modified black hole horizons exhibit parameter-dependent behavior} or modified gravity models \citep{Duan:2023gng}.

The allowed region of the parameter space is illustrated in Fig.~\ref{parameter} in the $(\ell,a/M)$ plane for different fixed values of the electric charge $Q$. The region below  red curve corresponds to configurations admitting horizons, while the region above represents solutions without a black hole structure. For a given Lorentz-violating parameter $\ell$, increasing the electric charge $Q$ reduces the maximum allowed value of the rotation parameter. Similarly, for fixed $Q$, an increase in the Lorentz-violating parameter $\ell$ leads to a decrease in the permitted range of the spin parameter. Therefore, both $Q$ and $\ell$ actions shrink the domain of parameters supporting regular black hole configurations compared to the Kerr case \citep{Kerr:1963ud, Duan:2023gng, Sucu:2025lqa}. The event horizon radius reduces monotonically as the spin parameter grows until it approaches a critical value of $a/M$, where the spacetime reaches an extremal with the Cauchy and event horizons coinciding \citep{Bardeen:1972fi}. Beyond this cutoff, the solution corresponds to a horizonless compact object and there are no actual positive roots of $\Delta(r)=0$.

To ensure the physical relevance of the photon dynamics, we limit our analysis of null geodesics and black hole shadows in the next section to the parameter space where black hole solutions are well-defined \citep{Perlick:2021aok}.

\begin{figure}[t]
\begin{centering}
\includegraphics[scale=0.65]{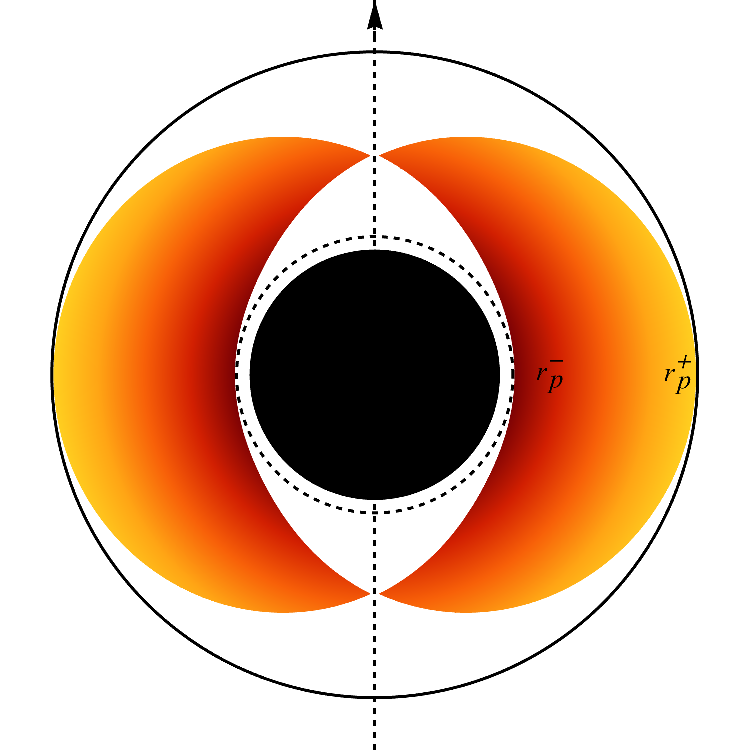}
\caption{Cross-section of the spherical photon region in the plane containing the spin axis for the charged rotating KR black hole with $a=0.8M$, $\ell=0.1$, and $Q=0.2M$. The shaded region corresponds to unstable spherical photon orbits. The black disk denotes the event horizon $r_{h}$, while the dashed and solid circular curves visualise the equatorial prograde $(r_{p}^{-})$ and retrograde $(r_{p}^{+})$ circular photon radii, respectively. The vertical dashed line indicates the rotation axis.}\label{pr1}	
\end{centering}
\end{figure}

\begin{figure*}
\centering
\begin{tabular}{c c}
\includegraphics[scale=0.65]{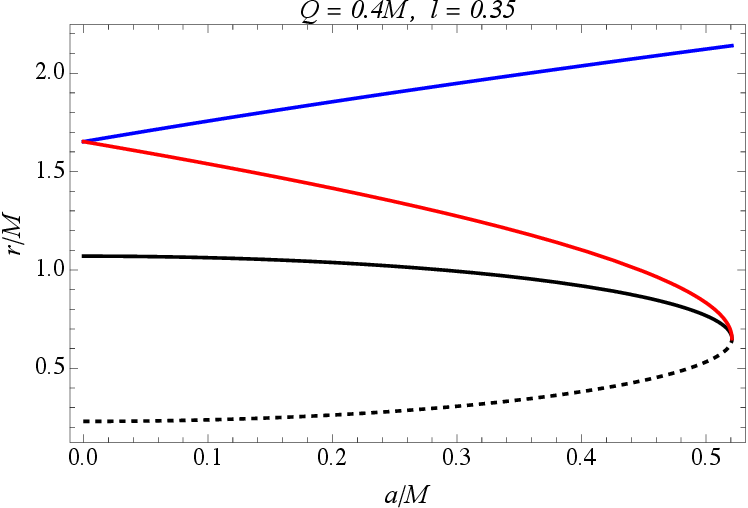} 
\hspace{1cm} &
\includegraphics[scale=0.65]{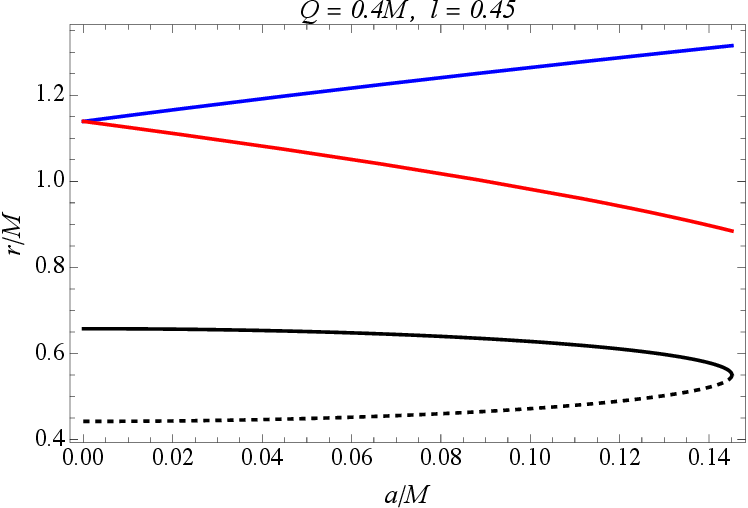}
\end{tabular}
\caption{The radial structure of the charged rotating KR black hole as a function of the spin parameter $a / M$. The solid black and dashed black curves represent the outer event horizon $(r_{+})$ and inner Cauchy horizon $(r_{-})$, respectively. The solid blue curve corresponds to the retrograde equatorial circular photon orbit $(r_{p}^{+})$, while the solid red curve denotes the prograde equatorial circular photon orbit $(r_{p}^{-})$.}
\label{fig:radii_vs_spin}
\end{figure*}

\section{Black hole shadow}\label{Sec-3}

The charged KR black hole spacetime is stationary and axisymmetric, and the corresponding null geodesics admit conserved quantities associated with these symmetries, namely the energy $\mathcal{E}$ and the axial angular momentum $\mathcal{L}$. 
Moreover, the Hamilton--Jacobi equation for geodesic motion is separable, implying the existence of a third constant of motion, known as the Carter constant \citep{Carter:1968rr, Chandrasekhar:1985kt, Afrin:2021wlj, Kumar:2020hgm}. 
As a consequence, the photon equations of motion can be expressed in first--order differential form as
\begin{align}
\Sigma \frac{dt}{d\tau} &=
\frac{r^2+a^2}{\Delta(r)}
\left(\mathcal{E}(r^2+a^2)-a\mathcal{L}\right)
- a\left(a\mathcal{E}\sin^2\theta-\mathcal{L}\right), \\
\Sigma \frac{d\phi}{d\tau} &=
\frac{a}{\Delta(r)}
\left(\mathcal{E}(r^2+a^2)-a\mathcal{L}\right)
-\left(a\mathcal{E}-\frac{\mathcal{L}}{\sin^2\theta}\right), \\
\Sigma \frac{dr}{d\tau} &= \pm\sqrt{\mathcal{R}(r)}, \\
\Sigma \frac{d\theta}{d\tau} &= \pm\sqrt{\Theta(\theta)}.
\end{align}
where $\tau$ is the affine parameter along the null geodesics and $\Sigma=r^2+a^2\cos^2\theta$. 
The functions $\mathcal{R}(r)$ and $\Theta(\theta)$, respectively, represent the radial and angular parts of the Hamilton--Jacobi equation and encode the full dynamics of photon motion \citep{Carter:1968rr, Chandrasekhar:1985kt}.
\begin{figure*}
    \begin{tabular}{c @{\hspace{2cm}} c}
     \includegraphics[scale=0.65]{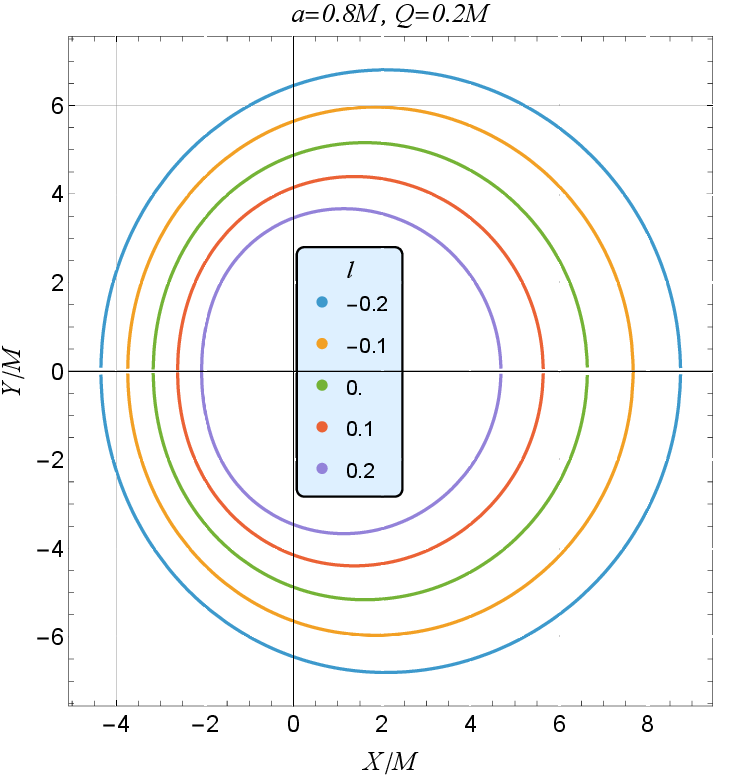}&
    \includegraphics[scale=0.65]{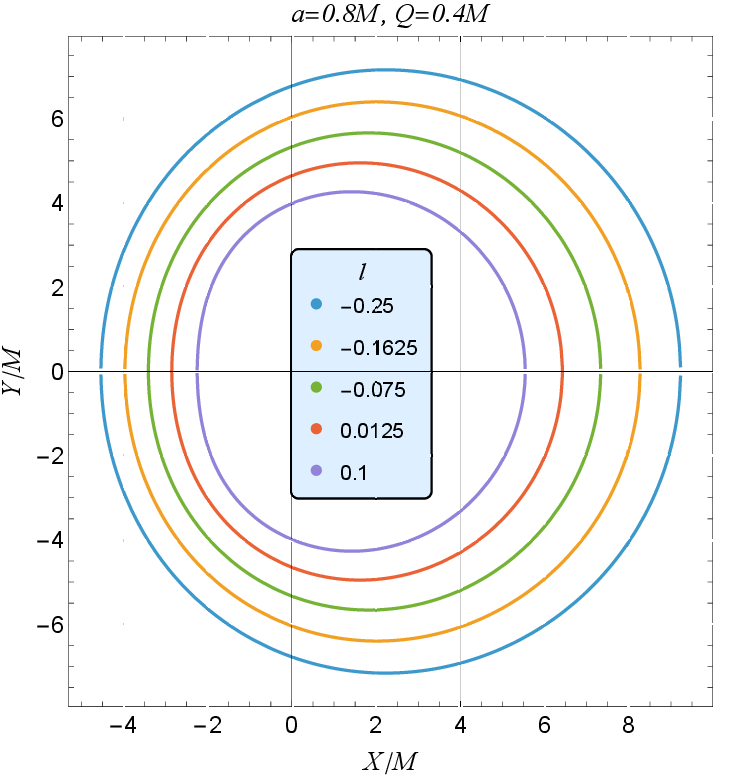}
    \end{tabular}
    \caption{Shadow geometry of the KR black hole for an equatorial observer $\theta_0 = \pi /2$. Left panel shows the effect of varying $\ell$ for fixed $(a, Q) = (0.8M, 0.2M)$, while the right panel illustrates the variation with $\ell$ for fixed $(a, Q) = (0.8M, 0.4M)$.}\label{fig2}
\end{figure*}
\begin{table*}
\centering
\caption{Equatorial circular prograde ($r_{P}^{-}$) and retrograde ($r_{P}^{+}$) photon orbit radii for the charged rotating KR black hole with fixed $\ell=0.1$ for $Q/M=0.2$ and $Q/M=0.4$, compared with the corresponding Kerr black hole values $r_{K}^{\mp}$ at different values of spin.}
\label{table1A}
\setlength{\tabcolsep}{8pt}
\renewcommand{\arraystretch}{1.4}
\begin{tabular}{||c|cc|cc|cc||}
\hline
 & \multicolumn{2}{c|}{Kerr Black Hole} & \multicolumn{2}{c|}{$Q/M=0.2$} & \multicolumn{2}{c||}{$Q/M=0.4$} \\
\hline
$a/M$ & $r_{K}^{+}/M$ & $r_{K}^{-}/M$ & $r_{P}^{+}/M$ & $r_{P}^{-}/M$ & $r_{P}^{+}/M$ & $r_{P}^{-}/M$ \\
\hline\hline
0.1 & 3.11335 & 2.88219 & 2.77506 & 2.55365 & 2.67293 & 2.44414 \\
0.2 & 3.22281 & 2.75919 & 2.87945 & 2.43515 & 2.78021 & 2.32076 \\
0.3 & 3.32885 & 2.63003 & 2.98037 & 2.31005 & 2.88361 & 2.18952 \\
0.4 & 3.43184 & 2.49336 & 3.07820 & 2.17672 & 2.98363 & 2.04819 \\
0.5 & 3.53209 & 2.34730 & 3.17327 & 2.03274 & 3.08064 & 1.89314 \\
0.6 & 3.62985 & 2.18891 & 3.26587 & 1.87413 & 3.17497 & 1.71783 \\
0.7 & 3.72535 & 2.01333 & 3.35621 & 1.69351 & 3.26687 & 1.50752 \\
0.8 & 3.81876 & 1.81109 & 3.44450 & 1.47340 & 3.35658 & 1.20629 \\
0.9 & 3.91027 & 1.55785 & 3.53090 & 1.13344 & - & - \\
\hline
\end{tabular}
\end{table*}

To analyze the radial behavior of photon trajectories, it is useful to express the radial motion in terms of an effective potential. For the effective potential analysis shown in Fig.~\ref{fig:Veff}, we restrict to the equatorial photon motion, corresponding to $\eta=0$, and set $\mathcal{E}=1$. In this case, the radial motion equation can be rewritten as
\begin{equation}
\left(\Sigma \frac{dr}{d\tau}\right)^2 + V_{\mathrm{eff}}(r) = 0,
\end{equation}
where
\begin{equation}
V_{\mathrm{eff}}(r)
=
-
\left[
\big((r^{2}+a^{2})-a\mathcal{L}\big)^{2}
-
\Delta(r)(\mathcal{L}-a)^{2}
\right].
\end{equation}
We have plotted $V_{\mathrm{eff}}(r)$ in Fig.~\ref{fig:Veff} for different values of the angular momentum $\mathcal{L}$ around its critical value $\mathcal{L}_c$. The turning points of the radial motion are determined by the condition $V_{\mathrm{eff}}(r)=0$, while unstable circular photon orbits correspond to local maxima of the effective potential \citep{Chandrasekhar:1985kt, Perlick:2021aok}.

Further, for null geodesics, it is convenient to introduce the following dimensionless impact parameters \citep{Chandrasekhar:1985kt, Afrin:2021wlj, Kumar:2020ltt, Kumar:2018ple}
\begin{equation}
\zeta \equiv \frac{\mathcal{L}}{\mathcal{E}}, 
\qquad
\eta \equiv \frac{\mathcal{K}}{\mathcal{E}^{2}} .
\end{equation}
Upon normalizing the radial equation by $\mathcal{E}^{2}$, the photon motion in the radial direction is governed by the radial function
\begin{equation}
\mathcal{R}(r)
=
\big[(r^{2}+a^{2})-a\zeta\big]^{2}
-
\Delta(r)\big[\eta+(a-\zeta)^{2}\big].
\label{eq19}
\end{equation}
The allowed region of photon motion is determined by the condition $\mathcal{R}(r)\geq 0$. 
The function $\mathcal{R}(r)$ completely characterizes the radial dynamics and determines the photon region 
that gives rise to the black hole shadow \citep{Bardeen:1972fi, Perlick:2021aok}. 
The boundary of the black hole shadow is governed by unstable spherical photon orbits, which form the 
separatrix between photon trajectories that escape to infinity and those captured by the black hole. 
These orbits correspond to null geodesics confined at a constant radial coordinate $r=r_p$, satisfying the conditions

\begin{equation}
\mathcal{R}(r_p)=0, \qquad
\left.\frac{d\mathcal{R}}{dr}\right|_{r=r_p}=0, \qquad
\left.\frac{d^{2}\mathcal{R}}{dr^{2}}\right|_{r=r_p}<0,
\label{eq20}
\end{equation}
where $r_p$ denotes the radius of the spherical photon orbit. The first two conditions ensure that the radial motion is stationary, while the third condition guarantees the instability of the orbit \citep{Chandrasekhar:1985kt, Bardeen:1972fi}.  The above conditions determine the critical values of the impact parameters $\zeta$ and $\eta$, denoted by $(\zeta_c,\eta_c)$, which correspond to unstable spherical photon orbits and define the boundary of the black hole shadow \citep{Perlick:2021aok}. For the charged KR black hole spacetime, solving Eq.~(\ref{eq20}) simultaneously yields $\zeta_c$ and $\eta_c$ as functions of the photon orbit radius $r_p$ and the black hole parameters $(M,a,\ell,Q)$:
\begin{equation}
\zeta_c
=
\frac{(r_p^2+a^2)\,\Delta'(r_p)-4r_p\,\Delta(r_p)}
     {a\,\Delta'(r_p)},
\label{eq21}
\end{equation}
\begin{equation}
\eta_c
=
\frac{
16a^2 r_p^2\,\Delta(r_p)
-
\left[(r_p^2+a^2)\,\Delta'(r_p)-4r_p\,\Delta(r_p)\right]^2
}
{a^2\,\Delta'(r_p)^2}.
\label{eq22}
\end{equation}
These expressions generalize the corresponding Kerr results to the case of a Lorentz-violating KR background, with deviations encoded through the modified function $\Delta(r)$ \citep{Kerr:1963ud, Duan:2023gng, Sucu:2025lqa}.

The unstable spherical photon orbits determined by Eq.~(\ref{eq20}) form a three-dimensional hypersurface known as the photon region. 
In rotating spacetimes, this region is bounded in the equatorial plane by the prograde and retrograde circular photon orbits and plays a central role in determining the structure of the black hole shadow \citep{Bardeen:1972fi, Perlick:2021aok}. 

In the limit ($\ell \to 0$, $Q \to 0$), the radii of the prograde ($r_{K}^{-}$) and retrograde ($r_{K}^{+}$) circular photon orbits in the equatorial plane admit exact analytical expressions \citep{Bardeen:1972fi, Chandrasekhar:1985kt}:
\begin{eqnarray}
r_{K}^{-} &=& 2M \left[ 1 + \cos \left( \frac{2}{3} \arccos \left( -\frac{|a|}{M} \right) \right) \right], \\
r_{K}^{+} &=& 2M \left[ 1 + \cos \left( \frac{2}{3} \arccos \left( \frac{|a|}{M} \right) \right) \right],
\end{eqnarray}
with the ranges $M \le r_{K}^{-} \le 3M$ and $3M \le r_{K}^{+} \le 4M$. 
We present these exact Kerr values in Table~\ref{table1A} as a baseline for comparison with our numerical results for the charged KR geometry.

To visualize the geometric structure in the charged rotating KR spacetime, we construct the photon region for representative parameter values, as shown in Fig.~\ref{figPRP}. 
The cross-sectional view illustrates the separation between the event horizon and the photon region.  As the spin increases, frame-dragging effects shift the prograde circular photon orbit inward and the retrograde orbit outward, producing an asymmetric thickening of the photon shell while remaining entirely outside the event horizon \citep{Chandrasekhar:1985kt, Perlick:2021aok}.

A cross-sectional view of the photon region in the plane containing the spin axis is depicted in Fig.~\ref{pr1}, clearly showing the spatial separation between the event horizon and the prograde and retrograde unstable photon orbits. 
To further quantify this radial structure in the KR black hole spacetime, we plot the characteristic radii as functions of the dimensionless spin parameter $a/M$ in Fig.~\ref{fig:radii_vs_spin} for fixed values $Q=0.4M$ and different values of $\ell$. 
The solid and dashed black curves represent the outer ($r_+$) and inner ($r_-$) horizons, respectively, which approach each other and merge at the extremal spin limit. 
The solid blue and red curves denote the retrograde ($r_p^{+}$) and prograde ($r_p^{-}$) equatorial circular photon orbits. 
Due to frame dragging, the prograde photon orbit shifts toward the event horizon with increasing spin, whereas the retrograde orbit moves outward, enhancing the asymmetry of the photon region \citep{Bardeen:1972fi}. It is important to note that for smaller values of $\ell$ and the corresponding extremal spin parameter $a = a_E$, the event horizon radius coincides with that of the prograde photon orbit. As $\ell$ increases, however, this coincidence breaks down, resulting in a finite separation between the two radii (cf. Fig.~\ref{fig:radii_vs_spin}). 

The apparent shape of the black hole shadow, as observed by a distant observer, is  described in terms of the celestial coordinates $(X,Y)$, which represent the apparent perpendicular distances of the photon image from the axis of symmetry and the equatorial plane, respectively. 
For an observer located at spatial infinity with inclination angle $\theta_0$, the celestial coordinates are given by \citep{Bardeen:1973tla, Chandrasekhar:1985kt, Hioki:2009na}
\begin{equation}
X = -\,\zeta_c \csc\theta_0 ,
\label{eq23}
\end{equation}
\begin{equation}
Y = \pm \sqrt{\eta_c + a^2\cos^2\theta_0 - \zeta_c^{\,2}\cot^2\theta_0 }.
\label{eq24}
\end{equation}

For an observer in the equatorial plane ($\theta_0=\pi/2$), Eqs.~(\ref{eq23}) and (\ref{eq24}) reduce to
\begin{equation}
\{X,Y\}=\{-\zeta_c,\pm\sqrt{\eta_c}\}.
\label{eq25}
\end{equation}
For the charged KR black hole metric given in Eq.~(\ref{28}), the celestial coordinates of the shadow for an equatorial observer take the explicit form
\begin{equation}
X=-\frac{(r_p^2+a^2)\,\Delta'(r_p)-4r_p\,\Delta(r_p)}
       {a\,\Delta'(r_p)},
\label{eq26}
\end{equation}
\begin{equation}
Y = \pm \sqrt{
\frac{
16a^2 r_p^2 \Delta(r_p)
-
\left[(r_p^2+a^2)\Delta'(r_p)-4r_p\Delta(r_p)\right]^2
}
{a^2 \Delta'(r_p)^2}
}.
\label{eq27}
\end{equation}
In the limit $\ell \to 0$, the $\Delta(r)$ reduces to that of the classical Kerr black hole, and Eqs.~(\ref{eq26})--(\ref{eq27}) simplify to
\begin{eqnarray}
X_{K} &=&
\frac{a^{2}(r_{p}+M)+r_{p}^{2}(r_{p}-3M)}{a(M-r_{p})}, \nonumber\\
Y_{K} &=&
\frac{r_{p}^{3/2}\left[4a^{2}M-r_{p}(r_{p}-3M)^{2}\right]^{1/2}}
{a(M-r_{p})},
\label{eq28}
\end{eqnarray}
which coincide with the well-known celestial coordinates of the Kerr black hole shadow \citep{Bardeen:1973tla, Hioki:2009na}, thereby consistently recovering the Kerr limit.

Using the celestial coordinates $(X,Y)$ derived above, we visualize the apparent shape of the photon ring as seen by a distant observer. In Fig.~\ref{fig2}, we compare the shadow contours for an equatorial observer across different values of the Lorentz-violating parameter, keeping the spin parameter fixed at $a = 0.8M$ and considering electric charges $Q = 0.2M$ and $Q = 0.4M$. For values below the extremal limit ($\ell_E \approx 0.2823$ for $Q = 0.2M$ and $\ell_E \approx 0.1424$ for $Q = 0.4M$) at $a/M = 0.8$, the configurations correspond to regular black holes.  In this regime, the photon rings form closed and distorted silhouettes, which are displaced horizontally and become increasingly flattened on the prograde side due to frame-dragging effects \citep{Bardeen:1973tla, Chandrasekhar:1985kt}. 

\subsection{No-Horizon Spacetime Shadow}
The parameter space of charged rotating KR black holes is illustrated in Fig.~\ref{parameter}, which shows the $(a/M, \ell)$ plane for fixed electric charges $Q = 0.2M$ (left panel) and $Q = 0.4M$ (right panel). The red curve outlines the boundary between black hole spacetimes and no-horizon configurations or naked singularities. Below this red curve, in a grey shaded region, spacetime possesses both a Cauchy and an event horizon -- black hole solutions. Above the red curve, no real positive roots of $\Delta(r) = 0$ exist, indicating horizonless spacetimes.

\begin{figure*}
\centering
\begin{tabular}{c c}
\includegraphics[scale=0.65]{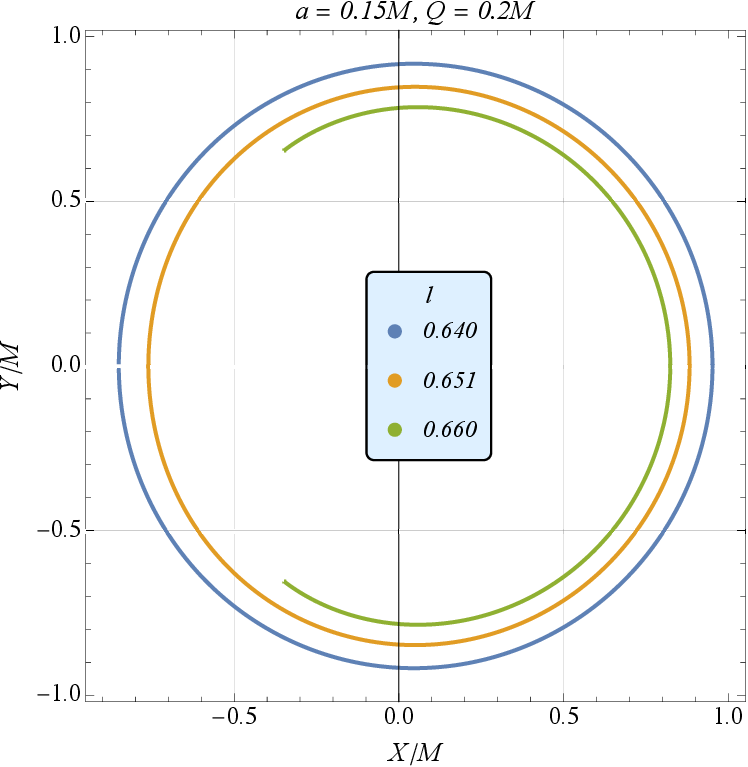} 
\hspace{1cm} &
\includegraphics[scale=0.65]{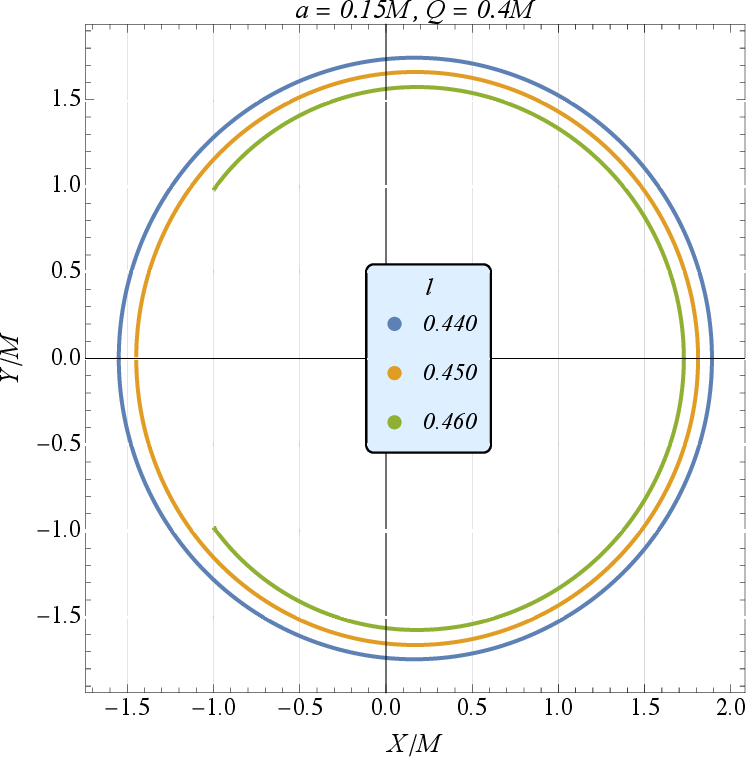}
\end{tabular}
\caption{Apparent photon rings of the charged rotating KR spacetime for an equatorial observer $\theta_0 = \pi /2$ in celestial coordinates $(X / M, Y / M)$, shown for fixed spin $a = 0.15M$. The left panel corresponds to $Q = 0.2M$, while the right panel corresponds to $Q = 0.4M$. In each panel, the blue curve represents a black hole configuration $(\ell < \ell_h)$, the yellow curve corresponds to a no-horizon spacetime that still admits a closed photon ring $(\ell_h < \ell < \ell_p)$, and the green curve represents a no-horizon spacetime with an open, arc-like photon ring $(\ell > \ell_p)$.}
\label{fig:shadow_spin}
\end{figure*}

A notable feature within the no-horizon regime is the black shaded region, which represents a subset of parameter space where closed photon rings exist despite the absence of an event horizon. In the black shaded region, the cumulative effects of rotation, electric charge, and Lorentz violation create a potential barrier sufficient to confine photons in closed orbits, thereby yielding a complete shadow  for the observer,  similar to that of a black hole. The unshaded region above the red curve represents horizonless spacetimes that do not admit closed photon rings, where the observable structure would consist of open, arc-like features rather than a closed shadow  Fig.~\ref{fig:shadow_spin}

The red boundary shifts downward as the electric charge increases, indicating that both $Q$ and the Lorentz-violating parameter $\ell$ reduce the parameter space supporting black hole configurations compared to the Kerr case. For a fixed $\ell$, increasing $Q$ lowers the maximum allowed spin $a$ for horizon existence; similarly, for fixed $Q$, increasing $\ell$ reduces the permissible spin range. The existence of closed photon rings in horizonless spacetimes (black shaded region) has important implications for interpreting EHT observations, as such configurations could potentially mimic black hole shadows while lacking an event horizon. Future high-resolution observations may be capable of distinguishing between these scenarios by probing the detailed morphology of the photon ring and the nature of the central dark region.

Table~\ref{tabNH} lists the critical values of the Lorentz-violating parameter $\ell$ that characterize the no-horizon regime for charged rotating KR spacetimes at fixed spin $a = 0.15M$. For a given  $Q$, the parameters $\ell_h$ and $\ell_p$ denote, respectively, the extremal boundary where the event horizon disappears and the upper limit beyond which closed photon rings cease to exist. When $\ell_h < \ell < \ell_p$, the horizonless spacetime admits a closed photon ring structure that can produce a black hole like shadow boundary, in contrast to the Kerr naked singularity, which yields only an open, arc-like photon ring. Increasing the charge $Q/M$ from $0.2$ to $0.4$ reduces both $\ell_h$ and $\ell_p$, narrowing the parameter window for closed photon ring formation. 

In such case, the behavior of photon orbits changes qualitatively: as shown in studies of rotating spacetimes \citep{Kumar:2020ltt, Kumar:2018ple, Perlick:2021aok}, prograde photon trajectories no longer form closed spherical orbits but instead spiral inward. 
Consequently, only retrograde photon trajectories contribute to the observable structure, producing an open, arc-like feature rather than a closed shadow boundary.

\begingroup
\begin{table}[t]
\centering
\setlength{\tabcolsep}{10pt} 
\begin{tabular}{||c|c|c||}
\hline
$Q/M$ & $\ell_h$ & $\ell_p$ \\
\hline\hline
0.2 & 0.650338 & 0.652821 \\
 0.4 & 0.449512 & 0.455741 \\
\hline
\end{tabular}
\caption{Values of $\ell$ in the no-horizon region at $a = 0.15M$. A closed photon ring forms only for $\ell_h < \ell < \ell_p$; while for $\ell > \ell_p$ the photon ring becomes open or arc-like.}
\label{tabNH}
\end{table}
\endgroup

\begin{figure}[!t]
\begin{center}
	\includegraphics[scale=0.85]{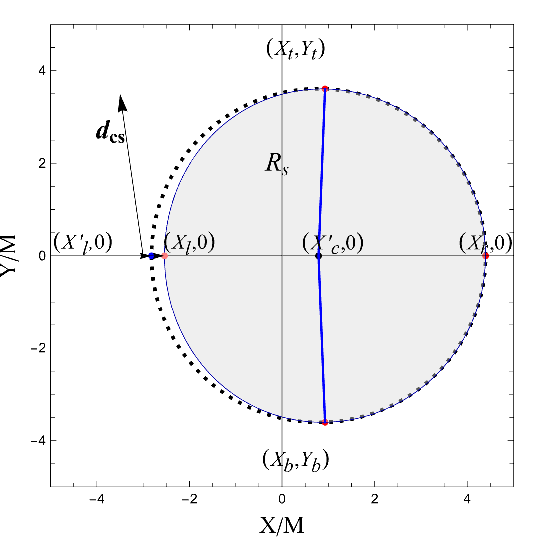}
	\caption{Geometric construction of the shadow observables for the charged rotating KR black hole. The shadow radius $R_s$ and the distortion parameter $\delta_s = d_{cs}/R_s$ are defined using a reference circle passing through three characteristic points of the shadow silhouette. The configuration shown corresponds to $a = 0.6M$, $\ell = 0.2$, $Q = 0.3M$, and $\theta_0 = \pi/2$.}\label{dis}
\end{center}
\end{figure}

\begin{figure*}[t]
\centering
\includegraphics[width=0.48\textwidth]{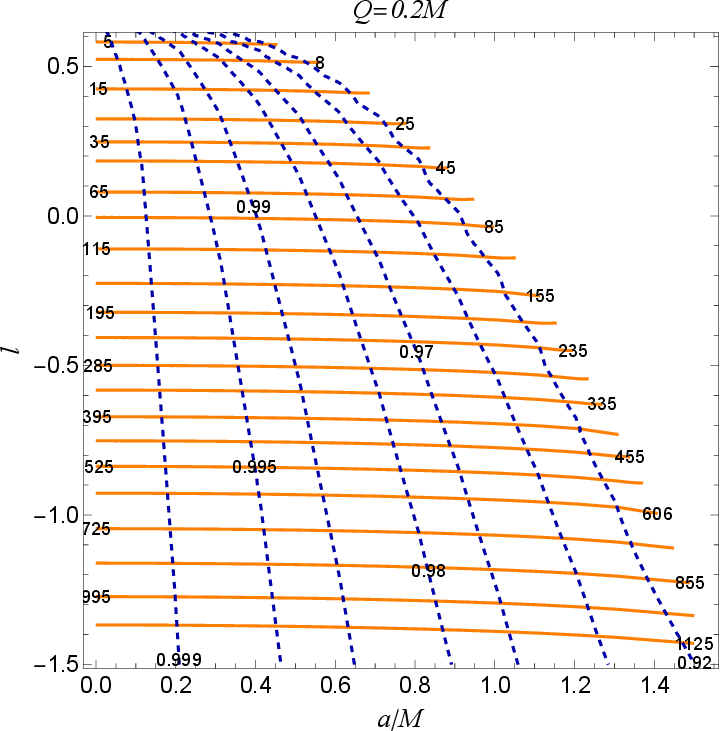}
\hfill
\includegraphics[width=0.48\textwidth]{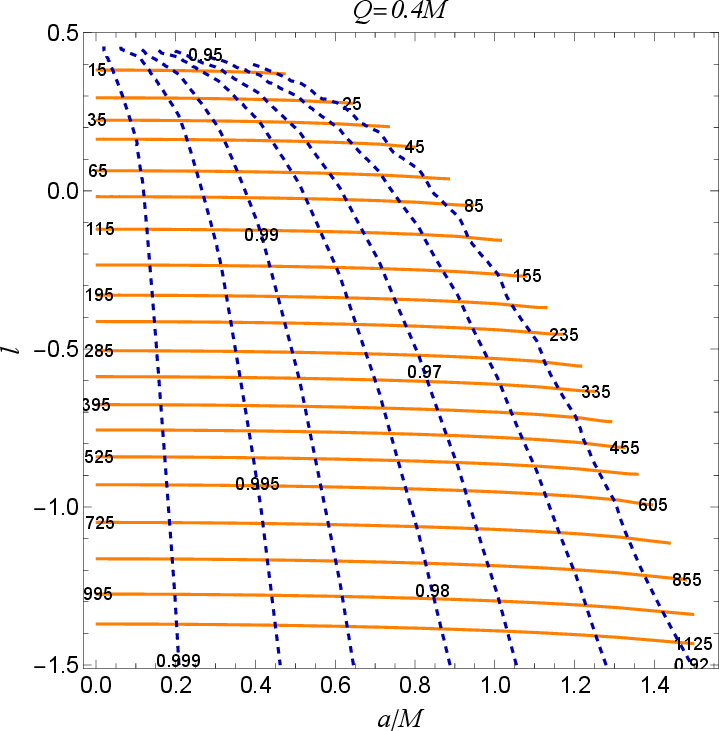}
\caption{
Contour plots of the shadow observables $A/M^2$ (red solid lines), representing the dimensionless shadow area, and $D$ (blue dashed lines), characterizing the shadow distortion, in the $(a,\ell)$ parameter space. The intersection of the $A/M^2$ and $D$ contours determines the corresponding values of $a$ and $\ell$. The left panel corresponds to $Q=0.2M$, while the right panel corresponds to $Q=0.4M$.
}
\label{contour2}
\end{figure*}

\begin{figure}
\centering
\includegraphics[scale=0.65]{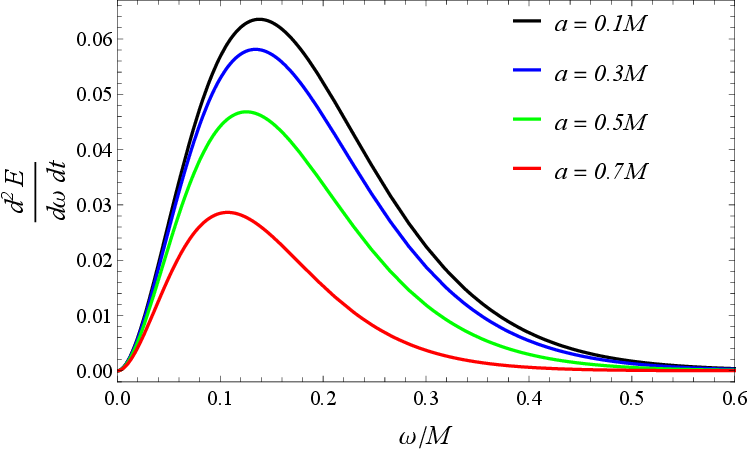}
\caption{
Energy emission rate $\frac{d^2 E}{d\omega dt}$ versus $\omega /M$ for different values of $a/M$, with $Q = 0.2M$ and $\ell = 0.1$.}
\label{fig:emission_a}
\end{figure}

\section{Parameter Estimation of black holes}\label{Sec-4}
Black hole parameter estimation is a robust method for analyzing deviations from the Kerr black hole in the strong-field regime. The no-hair theorem \cite{Carter:1968rr,Chandrasekhar:1985kt} states that in the context of general relativity, the properties of stationary black holes are entirely described by their mass, spin, and electric charge. So, any deviation from the Kerr geometry that can be observed could mean that there are more gravitational degrees of freedom or that the basic ideas of gravitational theory need to be changed.
In situations where Lorentz symmetry is violated or additional fields are present, these alterations can become apparent through changes in photon paths near the event horizon. The characteristics of the black hole shadow, a crucial tool for probing the fundamental spacetime structure \cite{Synge:1966,Bardeen:1973tla,Luminet:1979}, are thus affected.
Recent  EHT observation of horizon-scale images of M87* and Sagittarius A*, have enabled direct probes of the shadow structure of supermassive black holes \cite{EventHorizonTelescope:2019dse,EventHorizonTelescope:2019pgp, EventHorizonTelescope:2022xnr}. These observations make it possible to compare theoretical predictions with observational data in a meaningful way, especially shadow observables like angular diameter, asymmetry, and distortion. In this context, shadow-based parameter estimation provides a robust and largely model-independent approach to testing deviations from the Kerr hypothesis \cite{Johannsen:2010ru,Bambi:2019tjh,Vagnozzi:2022moj}.
The method developed by Hioki and Maeda \cite{Hioki:2009na}, which quantifies the shadow using two important observables—the distortion parameter $\delta_s$ and the shadow radius $R_s$—is a popular technique for describing the shadow geometry. By fitting a reference circle, as illustrated in Fig.~\ref{dis}, the observable $R_s$ determines the shadow's approximate size, and $\delta_s$ measures how far the shadow border deviates from circularity. This method provides a simple and intuitive way to extract black hole parameters, and has been extensively applied to a wide class of rotating and non-Kerr spacetimes \cite{Wei:2013kza,Abdujabbarov:2015xqa,Tsukamoto:2014tja,Cunha:2015yba,Younsi:2016azx,Kumar:2020owy,Afrin:2021imp}.
However, the Hioki--Maeda observables are based on a limited characterization of the shadow and may therefore fail to capture more subtle deformations that can arise in modified gravity scenarios. To address this shortcoming, a more refined framework was later introduced by Ghosh and Kumar \cite{Kumar:2018ple,Kumar:2020owy}, in which additional shadow observables were proposed to provide a more complete geometric characterization. In particular, this formalism includes higher-order shape descriptors and curvature-based quantities, thereby allowing a more accurate quantification of deviations from circularity.
The key advantages of the Ghosh–Kumar \cite{Kumar:2018ple,Kumar:2020owy} method can be summarized as follows. First, it captures detailed geometric features of the shadow beyond simple size and distortion, thereby improving sensitivity to small deviations from the Kerr geometry. Second, it reduces degeneracies between model parameters, especially in cases where different parameter combinations yield similar values of $(R_s,\delta_s)$. Third, it offers a more reliable framework for contrasting high-resolution observational data, such as those from the EHT, with theoretical expectations. Lastly, the approach is well-suited to systematic parameter estimation in non-Kerr spacetimes, where changes in gravity or the presence of extra fields may result in complex shadow morphologies.
\begin{table}[h]
\centering
\caption{Estimated values of parameters from observables $A$ and $D$ for different $Q$.}
\setlength{\tabcolsep}{4.5pt}
\renewcommand{\arraystretch}{1.15}
\setlength{\arrayrulewidth}{0.5pt}
\begin{tabular}{||c c| c c| c c||}
\hline
\multirow{2}{*}{$A/M^2$} & \multirow{2}{*}{$D$} 
& \multicolumn{2}{c}{$Q=0.2M$} 
& \multicolumn{2}{c||}{$Q=0.4M$} \\
\cline{3-4}
\cline{5-6}
 &  & $a/M$ & $\ell$ & $a/M$ & $\ell$ \\
\hline\hline
35  & 0.97  & 0.5564 & 0.2386 & 0.4887 & 0.2104 \\
45  & 0.999 & 0.1151 & 0.1893 & 0.1093 & 0.1567 \\
65  & 0.999 & 0.1251 & 0.0738 & 0.1126 & 0.6083 \\
65  & 0.97  & 0.6317 & 0.0684 & 0.5896 & 0.0579 \\
85  & 0.98  & 0.5597 & -0.0192 & 0.5354 & -0.0245 \\
115 & 0.97  & 0.7011 & -0.1236 & 0.6755 & -0.1346 \\
155 & 0.99  & 0.3262 & -0.2299 & 0.44   & -0.241 \\
195 & 0.98  & 0.6451 & -0.3316 & 0.6347 & -0.343 \\
\hline
\end{tabular}
\label{tab:parameter_estimation}
\end{table}
We perform parameter estimation for the charged, rotating black hole solution in the presence of a KR field, characterized by the parameter set $(M, a, \ell, Q)$. The black hole mass $M$ is fixed using observational priors, while the spin parameter $a$ and the Lorentz-violating parameter $\ell$ are treated as the primary parameters to be constrained.  The shadow observables, computed using the Ghosh–Kumar techniques, are then employed to quantify deviations from the Kerr geometry and constrain the underlying model parameters.

The dependence of the shadows on the spin $a$, the Lorentz-violating parameter $\ell$, and the observer inclination $\theta_0$ has been thoroughly characterized in the preceding section. Next,  shadow observables are employed to determine the allowed regions in the parameter space via contour analysis, extending the geometric parameter estimation originally developed in Refs.~\citep{Kumar:2018ple,Hioki:2009na}. 

The estimation scheme of Kumar \& Ghosh \citep{Kumar:2018ple} uses two robust observables — the \textbf{shadow area} $A$ and the \textbf{oblateness (axis ratio)} $D$ — which capture the overall size and deviation from circularity of the shadow, respectively. These are defined as
\begin{align}
A &= 2\int_{r_p^-}^{r_p^+} Y(r_p)\,\frac{dX(r_p)}{dr_p}\,dr_p, \\
D &= \frac{X_r-X_l}{Y_t-Y_b},
\end{align}
where $r_p^\pm$ denote the radii of the retrograde and prograde unstable spherical photon orbits that delineate the shadow boundary in the observer’s celestial plane~\citep{Kumar:2018ple}. For a spherically symmetric spacetime, $D=1$, whereas deviations from unity quantify deformation induced by rotation and additional spacetime structure, as discussed in Refs.~\citep{Grenzebach:2014fha, Abdujabbarov:2015xqa}.  
Contour curves of constant $A$ and $D$ in the $(a/M,\ell)$ parameter plane are shown in Fig.~\ref{contour2} for fixed $Q=0.2M$ (left) and $Q=0.4M$ (right). Their intersections estimate an independent  $(a/M,\ell)$, analogous to contour intersection techniques employed for other Kerr-like and hairy black hole spacetimes~\citep{Kumar:2018ple, Afrin:2023uzo}. The qualitative agreement between the parameter regions inferred from the traditional radius–distortion pair $(R_s,\delta_s)$ \cite{Hioki:2009na}.  and those obtained from $(A,D)$ demonstrates the internal consistency of the shadow-based parameter estimation framework, as corroborated in Refs.~\citep{Hioki:2009na, Tsukamoto:2017fxq}.  

The combined use of multiple shadow observables is essential for breaking parameter degeneracies that would otherwise arise if only a single observable were considered~\citep{Kumar:2018ple}. While the electric charge $Q$ enters the radial photon potential and influences the shadows, its contribution to the overall shadow  remains subdominant compared to the effects of spin and the Lorentz-violating parameter. Consequently, the dominant constraints are driven by variations in $(a,\ell)$.  

 A salient feature of the present analysis is the contrasting behavior of the event horizon area and the shadow area under variations of the Lorentz-violating parameter. Increasing $\ell$ tends to reduce the event horizon area but concurrently enlarges the shadow area, highlighting that the apparent shadow size does not scale directly with the physical horizon size. This reflects how the KR field modifies the unstable photon orbits forming the shadow boundary — a geometric projection of null geodesics rather than a direct measure of the horizon~\citep{Grenzebach:2014fha, Abdujabbarov:2015xqa}.  

In the Kerr limit $\ell\to 0$ and the neutral limit $Q\to 0$, all shadow observables smoothly reduce to their well-known Kerr expressions, recovering the standard Kerr shadow in the appropriate parameter limit~\citep{Hioki:2009na}.
\begin{figure*}
\centering
\includegraphics[width=0.48\textwidth]{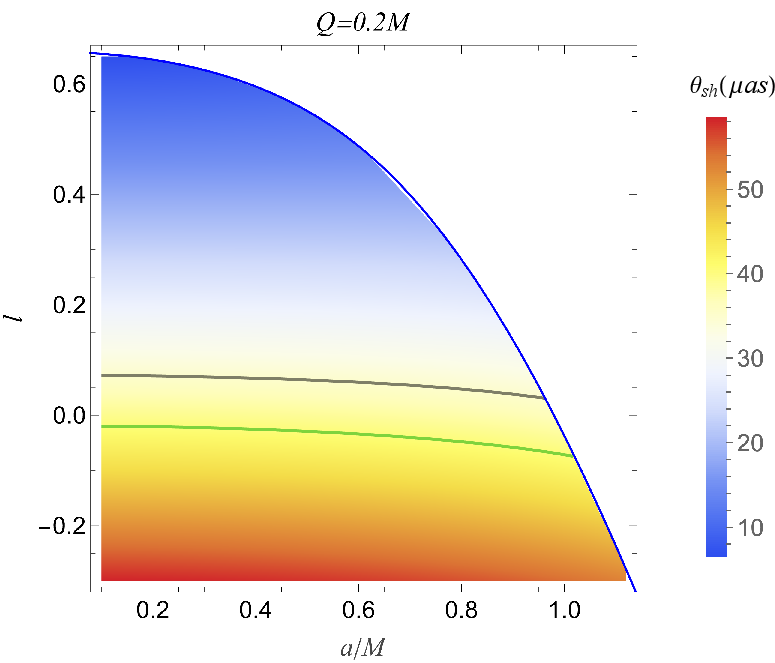}
\hfill
\includegraphics[width=0.48\textwidth]{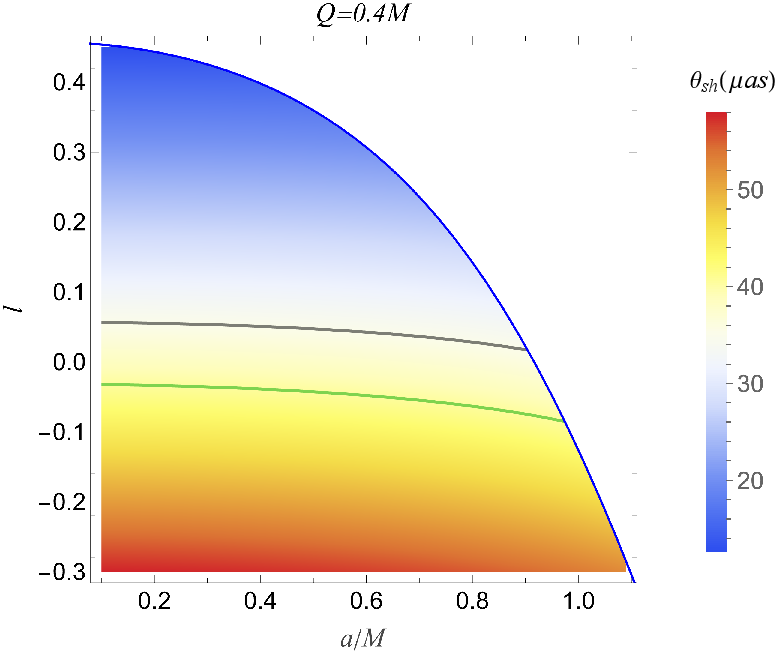}
\caption{EHT bounds on the shadow angular diameter $\theta_{sh}$ of the rotating KR black hole for M87* at $\theta_0=17^\circ$ . The allowed range $35.1\,\mu\mathrm{as} \lesssim \theta_{sh} \lesssim 40.5\,\mu\mathrm{as}$ represents the $1\sigma$ observational bounds. The blue curve denotes the extremal boundary, and the white region corresponds to the forbidden region.}
\label{M871}
\end{figure*}

\begin{figure*}
\begin{tabular}{c c}
\includegraphics[width=0.48\textwidth]{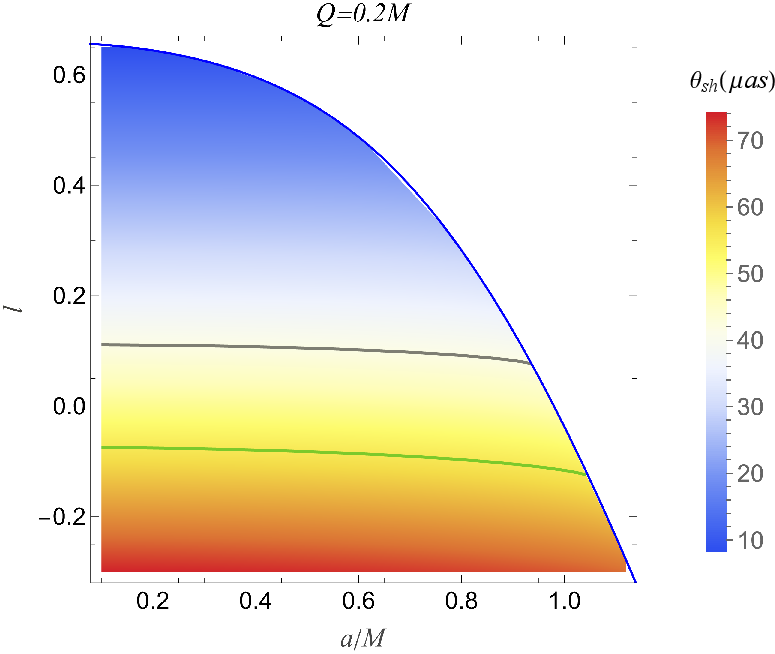}&
\hfill 
\includegraphics[width=0.48\textwidth]{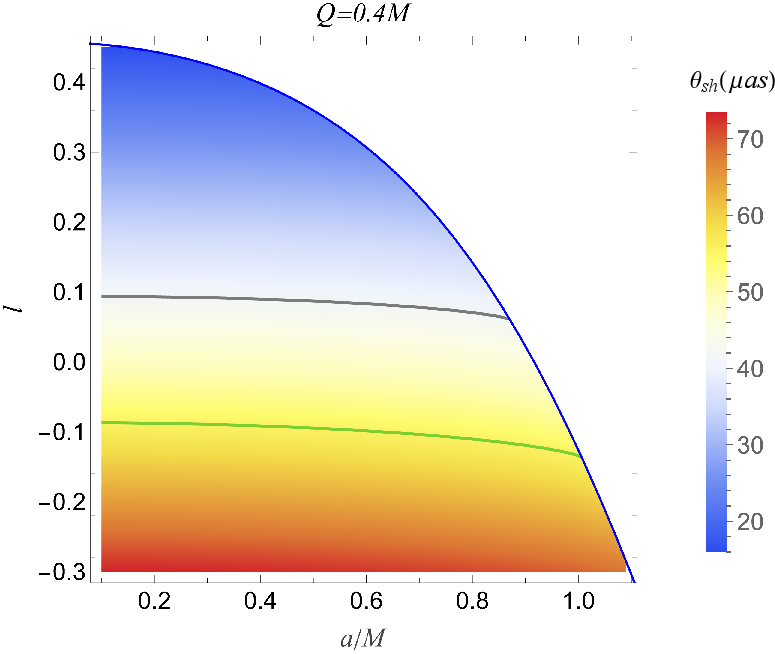}
\end{tabular}
\caption{EHT bounds on the shadow angular diameter $\theta_{sh}$ of the rotating KR black hole for Sgr~A* at $\theta_0=50^\circ$. The allowed range $41.7\,\mu\mathrm{as} \lesssim \theta_{sh} \lesssim 55.7\,\mu\mathrm{as}$ represents the $1\sigma$ observational bounds. The blue curve denotes the extremal boundary, and the white region corresponds to the forbidden region.}
\label{sgr1}
\end{figure*}

\subsection{Thermal Energy Emission}
The study of the energy emission rate provides an important complementary probe to shadow analysis, as it connects the geometric properties of null geodesics with the thermodynamic behaviour of black holes. While the shadow characterizes the photon capture region determined by unstable photon orbits, it is well known that, in the geometric optics (eikonal) limit, the high-energy absorption cross-section is also governed by the same photon region~\citep{Decanini:2011xi, Sanchez:1977vz}. Consequently, the shadow radius offers a direct estimate of the effective absorptive area of the black hole~\citep{Wei:2013kza}.

From a physical perspective, Hawking radiation endows black holes with a thermal spectrum~\citep{Hawking:1975vcx}, and the corresponding energy emission rate depends on both the absorption cross-section and the Hawking temperature. Therefore, analyzing the emission spectrum provides a natural bridge between observable shadow features and fundamental thermodynamic quantities. 

\begin{figure*}
\begin{tabular}{c c}
\includegraphics[width=0.48\textwidth]{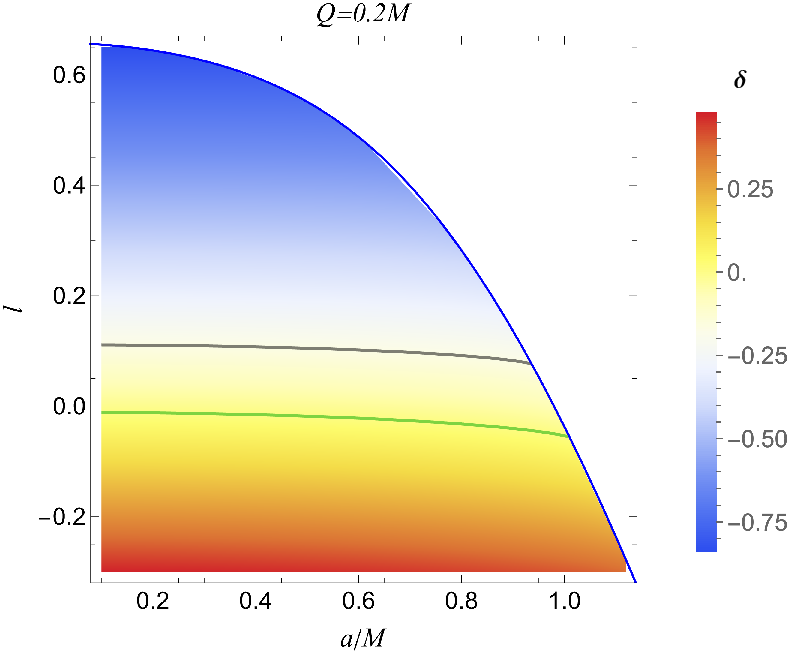}
\hfill &
\includegraphics[width=0.48\textwidth]{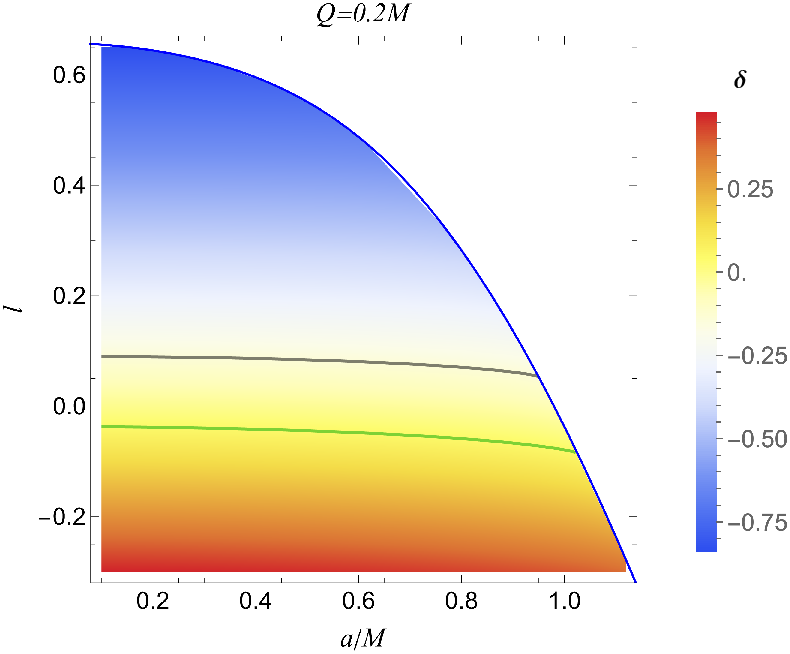}\\
\includegraphics[width=0.48\textwidth]{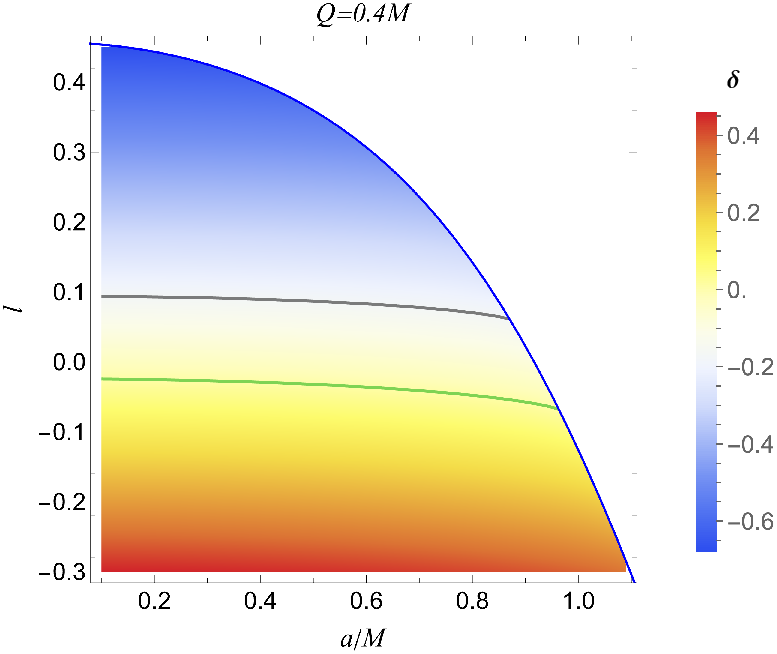}
\hfill &
\includegraphics[width=0.48\textwidth]{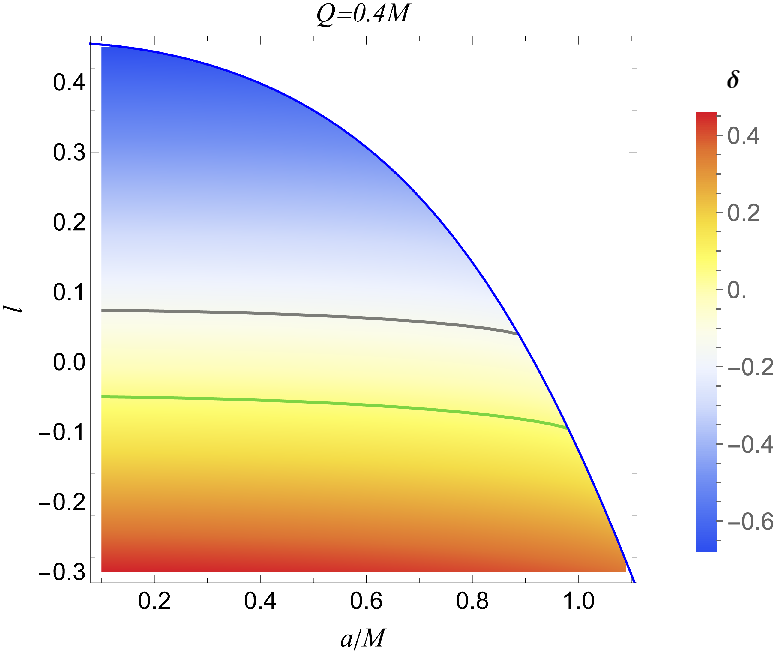}
\end{tabular}
\caption{Schwarzschild deviation for Sgr~A* in KR black hole model at an inclination angle of $\theta_0=50^\circ$. The contours correspond to observational bounds from VLTI ($\delta=-0.08^{+0.09}_{-0.09}$), on left, and Keck ($\delta=-0.04^{+0.09}_{-0.10}$), on right, measurements. The white region is forbidden region.}
\label{sgr2}
\end{figure*}

\begin{figure*}
    \includegraphics[width=0.48\textwidth]{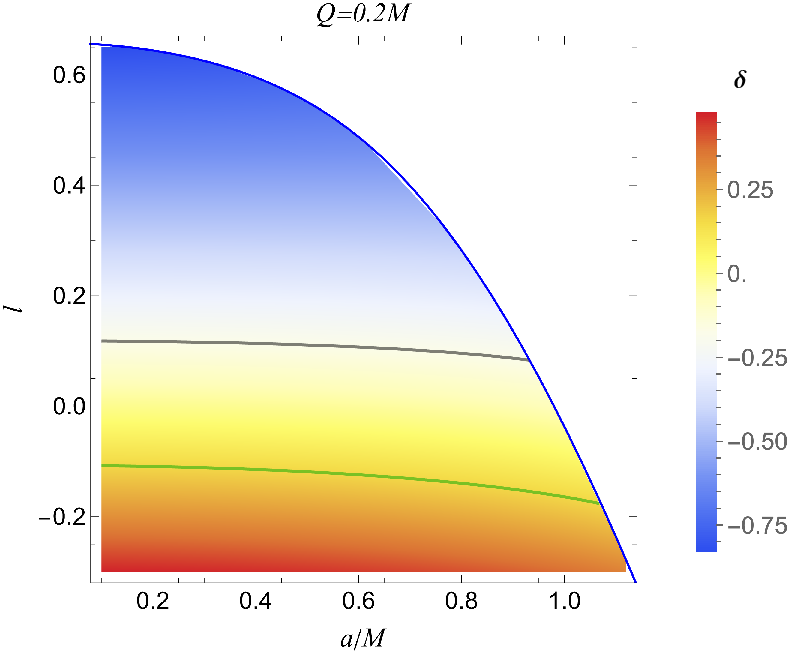}
\hfill
    \includegraphics[width=0.48\textwidth]{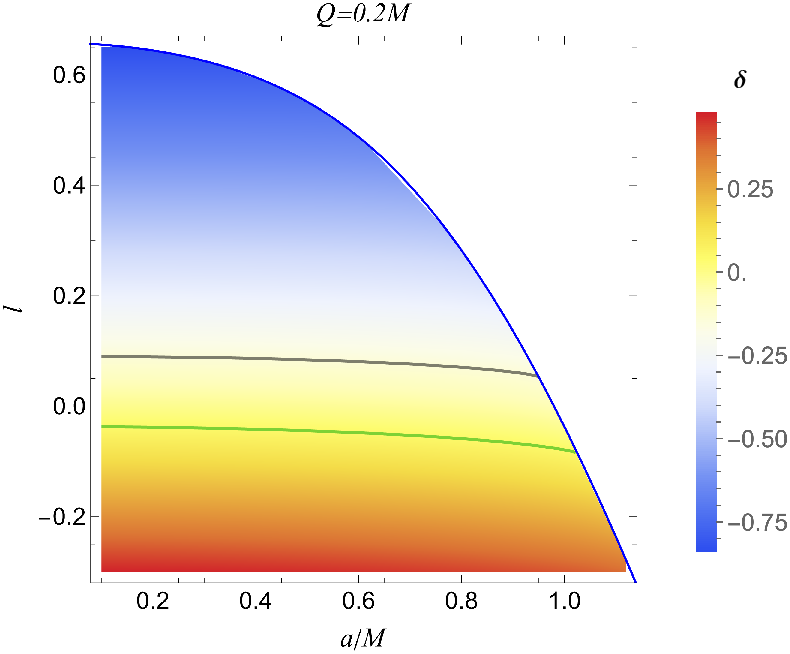}
    \caption{Schwarzschild deviation for M87* in KR black hole model at an inclination angle of $\theta_0=17^\circ$, corresponding to $\delta=-0.01^{+0.17}_{-0.17}$. The white region is forbidden region.}
    \label{M872}
\end{figure*}

The shadow radius provides a useful estimate of the high-energy absorption cross section of the charged, rotating KR black hole. In the geometric optics (eikonal) limit, the absorption cross-section for massless particles approaches a limiting constant value governed by the photon region, with small oscillatory corrections around it. Following~\citep{Wei:2013kza, Decanini:2011xi}, the high-energy absorption cross-section can be expressed as
\begin{equation}
\sigma_{\text{abs}} \approx \pi R_s^2 ,
\end{equation}
where $R_s$ is the effective shadow radius extracted from the celestial coordinates.  Assuming a thermal spectrum associated with Hawking radiation, the differential energy emission rate per unit frequency is given by
\begin{equation}
\frac{d^2E(\omega)}{d\omega dt}
=
\frac{\sigma_{\text{abs}}}{\pi^2}
\frac{\omega^3}{\exp\!\left(\omega/T_H\right)-1}.
\end{equation}
Substituting $\sigma_{\text{abs}} \approx \pi R_s^2$, one obtains
\begin{equation}
\frac{d^2E(\omega)}{d\omega dt}
=
\frac{R_s^2}{\pi}
\frac{\omega^3}{\exp\!\left(\omega/T_H\right)-1},
\label{emissionrate}
\end{equation}
which relates the emission spectrum directly to the shadow radius. This expression highlights that the shadow encodes the effective radiative area of the black hole in the high-frequency limit~\citep{Wei:2013kza}.

The Hawking temperature is determined from the surface gravity at the event horizon. For a stationary axisymmetric spacetime, it can be written as
\begin{equation}
T_H=
\frac{\Delta'(r_h)}{4\pi \left(r_h^2+a^2\right)},
\label{THexplicit}
\end{equation}
where $r_h$ is the largest real root of $\Delta(r_h)=0$. Since the horizon radius is determined implicitly through the metric function $\Delta(r)$, the temperature inherits a nontrivial dependence on the parameters $(a,\ell,Q)$.

It is worth emphasising that, although the shadow radius governs the high-energy absorption cross-section, the emission spectrum is primarily controlled by the Hawking temperature. Consequently, variations in the Lorentz-violating parameter $\ell$ influence the emission rate both directly through modifications of the horizon structure (affecting $T_H$) and indirectly via changes in the photon region (affecting $R_s$). This interplay encodes important information about the underlying spacetime geometry.

The emission spectrum is therefore controlled by both the shadow radius and the Hawking temperature. Increasing the spin parameter $a$ generally reduces the temperature through its influence on the horizon structure and the factor $(r_h^2+a^2)$, leading to a suppression of the emission peak. The Lorentz-violating parameter $\ell$ changes the spectrum by altering the horizon's position and the shape of $\Delta(r)$. This, in turn, affects both $T_H$ and $R_s$. Although the electric charge $Q$ does not appear explicitly in Eq.~(\ref{THexplicit}), it contributes indirectly through its effect on the horizon radius.

The resulting energy emission spectra, shown in Fig.~\ref{fig:emission_a}, exhibit a characteristic behaviour: the emission rate increases from zero at low frequencies, reaches a single peak around $\omega \sim T_H$, and then decays exponentially at higher frequencies.
\begin{table}[t]
\centering
\caption{EHT bounds on the Lorentz-violating parameter $\ell$ for M87* and Sgr A* at fixed charge $Q/M$. Allowed intervals follow from the angular diameter constraints $\theta_{\rm sh}\in[35.1,,40.5]~\mu{\rm as}$ (M87*) and $\theta_{\rm sh}\in[41.7,,55.7]~\mu{\rm as}$ (Sgr A*), as shown in Figs.~\ref{M871} and \ref{sgr1}.}
\label{tab:lbounds}
\renewcommand{\arraystretch}{1.2}
\setlength{\tabcolsep}{10pt}
\begin{tabular}{||c|ccc||}
\hline
$Q/M$ & BH & $a/M$ & $\ell$ bound \\
\hline\hline
\multirow{4}{*}{$0.2$}
& \multirow{2}{*}{M87*}   & $0.1001$ & $-0.019 \lesssim \ell \lesssim 0.075$ \\
&                         & $0.9583$ & $-0.076 \lesssim \ell \lesssim 0.029$ \\
\cline{2-4}
& \multirow{2}{*}{Sgr A*} & $0.1001$ & $-0.075 \lesssim \ell \lesssim 0.110$ \\
&                         & $0.923$  & $-0.124 \lesssim \ell \lesssim 0.076$ \\
\hline
\multirow{4}{*}{$0.4$}
& \multirow{2}{*}{M87*}   & $0.1001$ & $-0.031 \lesssim \ell \lesssim 0.053$ \\
&                         & $0.9013$ & $-0.073 \lesssim \ell \lesssim 0.017$ \\
\cline{2-4}
& \multirow{2}{*}{Sgr A*} & $0.1001$ & $-0.090 \lesssim \ell \lesssim 0.095$ \\
&                         & $0.861$  & $-0.110 \lesssim \ell \lesssim 0.065$ \\
\hline
\end{tabular}
\end{table}

\section{Constraints from the EHT Observation}\label{Sec-5}

The angular diameter of the black hole shadow provides a direct observable for comparison with EHT measurements of M87* and Sagittarius~A* \citep{EventHorizonTelescope:2019dse, EventHorizonTelescope:2022xnr, EventHorizonTelescope:2022xqj}. 
For M87*, the observed emission ring has an angular diameter of approximately $42\pm3\,\mu{\rm as}$, corresponding to a shadow diameter of $\theta_{\rm sh}=37.8\pm2.7\,\mu{\rm as}$. For Sagittarius~A*, the observed emission ring diameter is $51.8\pm2.3\,\mu{\rm as}$, with a corresponding shadow diameter of $\theta_{\rm sh}=48.7\pm7\,\mu{\rm as}$. These bounds correspond to the $1\sigma$ (68\% confidence) intervals reported by the EHT collaboration.

From the numerically determined shadow contour in celestial coordinates $(X,Y)$, the shadow area $A$ is computed using Green’s theorem, and the corresponding angular diameter is defined as
\begin{equation}
\theta_{\rm sh}=\frac{2}{d}\sqrt{\frac{A}{\pi}},
\label{angularDiameterEq}
\end{equation}
where $A$ is expressed in gravitational units $(GM/c^2)^2$ and $d$ denotes the source distance. Throughout this analysis, the electric charge is fixed to $Q=0.2M$, while the spin parameter $a$ and the Lorentz-violating parameter $\ell$ are varied.

To quantify departures from the Schwarzschild reference case, we introduce the fractional deviation parameter
\begin{equation}
\delta=
\frac{R_{\rm sh}}{3\sqrt{3}\,M}-1,
\end{equation}
where $R_{\rm sh}=\sqrt{A/\pi}$. In the Schwarzschild limit $R_{\rm sh}=3\sqrt{3}M$, giving $\delta=0$. Negative values of $\delta$ indicate a reduction in the shadow size relative to the Schwarzschild case, which is a common feature of rotating and non-Kerr geometries. For Kerr black holes, the deviation typically lies in the range $\delta\in[-0.075,0]$, providing a useful benchmark for assessing deviations from general relativity.

For M87*, we adopt $M=6.5\times10^9\,M_\odot$ and
$D=16.8\,{\rm Mpc}$ with inclination $\theta_0=17^\circ$.
Figure~\ref{M871} shows the angular diameter as a function of $(a/M,\ell)$. The contour lines at $\theta_{\rm sh}=35.1\,\mu{\rm as}$ and $\theta_{\rm sh}=40.5\,\mu{\rm as}$ represent the adopted $1\sigma$ bounds. The region between these contours is compatible with the EHT observations, while the white region corresponds to parameter values for which no horizon exists. The blue curve denotes the extremal boundary determined from the horizon condition. 

\begin{table}[t]
\centering
\caption{EHT bounds on $\ell$ from the Schwarzschild deviation parameter $\delta$ for Sgr~A* in the KR black hole model. Allowed intervals correspond to $\delta=-0.08^{+0.09}_{-0.09}$ (VLTI) and $\delta=-0.04^{+0.09}_{-0.10}$ (Keck), shown in Fig.~\ref{sgr2}.}
\label{tab:deltabounds}
\renewcommand{\arraystretch}{1.2}
\setlength{\tabcolsep}{10pt}
\begin{tabular}{||c|c|c||}
\hline
$Q/M$ & $a/M$ & $\ell$ bound \\
\hline\hline

\multicolumn{3}{||c||}{VLTI} \\
\hline
\multirow{2}{*}{$0.2$}
& $0.1001$ & $-0.0127\lesssim \ell \lesssim0.1125$ \\
& $0.9209$ & $-0.043\lesssim \ell \lesssim0.073$ \\
\hline
\multirow{2}{*}{$0.4$}
& $0.1001$ & $-0.0234\lesssim \ell \lesssim0.0957$ \\
& $0.8581$ & $-0.052\lesssim \ell \lesssim0.0625$ \\
\hline\hline

\multicolumn{3}{||c||}{Keck} \\
\hline
\multirow{2}{*}{$0.2$}
& $0.1001$ & $-0.0373\lesssim \ell \lesssim0.0916$ \\
& $0.9371$ & $-0.0711\lesssim \ell \lesssim0.0535$ \\
\hline
\multirow{2}{*}{$0.4$}
& $0.1001$ & $-0.049\lesssim \ell \lesssim0.0755$ \\
& $0.8762$ & $-0.0813\lesssim \ell \lesssim0.0381$ \\
\hline

\end{tabular}
\end{table}

\begin{table}[t]
\centering
\caption{EHT bounds on $\ell$ from the Schwarzschild deviation parameter $\delta$ for M87* in the KR black hole model. The allowed intervals correspond to $\delta=-0.01^{+0.17}_{-0.17}$, as shown in Fig.~\ref{M872}}
\label{tab:m87deltabounds}
\renewcommand{\arraystretch}{1.2}
\setlength{\tabcolsep}{10pt}
\begin{tabular}{||c|c|c||}
\hline
$Q/M$ & $a/M$ & $\ell$ bound \\
\hline\hline

\multirow{2}{*}{$0.2$}
& $0.1001$ & $-0.1077\lesssim \ell \lesssim0.1172$ \\
& $0.9240$ & $-0.1545\lesssim \ell \lesssim0.0806$ \\
\hline

\multirow{2}{*}{$0.4$}
& $0.1001$ & $-0.1225\lesssim \ell \lesssim0.0991$ \\
& $0.8582$ & $-0.1583\lesssim \ell \lesssim0.0651$ \\
\hline

\end{tabular}
\end{table}

For Sagittarius~A*, we use $M=4.0\times10^6\,M_\odot$,
$D=8.15\,{\rm kpc}$, and inclination $\theta_0=50^\circ$.
The angular diameter contours are shown in Fig.~\ref{sgr1}. The levels at $\theta_{\rm sh}=41.7\,\mu{\rm as}$ and $\theta_{\rm sh}=55.7\,\mu{\rm as}$ represent the observational bounds. The allowed parameter region lies between these contours. As in the M87* case, increasing $\ell$ leads to a reduction in the shadow size for fixed spin, shifting the allowed region accordingly.

A fully consistent comparison with non-Kerr geometries would require constructing dedicated GRMHD image libraries for the KR solution. However, existing studies indicate that current observational resolution does not significantly distinguish between Kerr and moderately deformed non-Kerr shadows. Differences in angular diameter are typically smaller than current measurement uncertainties, allowing the EHT bounds to be applied directly at the present level of precision.

Overall, the angular-diameter constraints restrict the admissible region in the $(a/M,\ell)$ parameter space for fixed $Q=0.2M$. Within this region, the charged rotating KR black hole remains consistent with current EHT observations of both M87* and Sgr~A*.

Table~\ref{tab:lbounds} summarizes the allowed ranges for the Lorentz-violating parameter $\ell$ obtained from the EHT angular diameter measurements of M87* and Sagittarius A*, for two different charge choices $Q/M = 0.2$ and $0.4$ shown in Figs.~\ref{M871} and \ref{sgr1}. A few things stand out. First, the bounds are not extremely tight---$\ell$ is typically constrained to within about $\pm0.1$ or so---but they are real, coming directly from the observed shadow sizes. Second, the constraints vary noticeably with both spin and charge. For M87* at the lower spin, $\ell$ lies between $-0.019$ and $0.075$ when $Q/M = 0.2$ for lower spin; cranking the spin around $0.95$ shifts this window downward to between $-0.076$ and $0.029$. Similar trends for Sgr A*: higher spin shifts the allowed $\ell$ range toward more negative values. Third, increasing the charge from $0.2$ to $0.4$ tightens the bounds somewhat and also shifts them, as seen most clearly in the M87* low-spin case where the upper bound drops from $0.075$ to $0.053$. These numbers tell us that if KR gravity is correct, the parameter $\ell$ cannot be arbitrarily large;  $\ell$  has to sit in a fairly narrow window that depends on the black hole's spin and charge.

Table~\ref{tab:deltabounds} and Fig.~\ref{sgr2} show what the Schwarzschild deviation parameter $\delta$ for Sgr A*—taken from VLTI and Keck measurements—implies for the Lorentz-violating parameter $\ell$ in the KR black hole model. A few things jump out. The VLTI constraints, with $\delta = -0.08^{+0.09}_{-0.09}$, give $\ell$ ranges that are generally tighter and sit a bit more toward positive values compared to what the Keck numbers yield. Take $Q/M = 0.2$ at low spin: VLTI allows $-0.0127 \lesssim \ell \lesssim 0.1125$, whereas Keck, with its less negative central value of $\delta = -0.04$, opens up a slightly wider window of $-0.0373 \lesssim \ell \lesssim 0.0916$ that dips further into negative territory. That makes sense—if the data want a less negative $\delta$, the model can afford a smaller $\ell$ to match.
Higher value spin narrows the allowed $\ell$  and tightens it toward more negative values, whether one looks at VLTI or Keck. For VLTI with $Q/M = 0.2$, at higher spin it tightens noticeably to $-0.043 \lesssim \ell \lesssim 0.073$. The same story plays out for $Q/M = 0.4$ and for the Keck data. For  $Q/M = 0.2$ to $0.4$ tends to shrink the $\ell$ window and shift it slightly lower, especially at higher spins. The Keck high-spin case illustrates this nicely: the upper bound drops from $0.0535$ to $0.0381$, while the lower bound moves from $-0.0711$ to $-0.0813$.
That consistency across two different observables gives some confidence that the numbers are not just a fluke of the method. For now, the KR black hole still fits the data just fine across a decent chunk of parameter space, though one suspects that future, sharper measurements will start chipping away at that window.

Table~\ref{tab:m87deltabounds} and Fig.~\ref{M872} show what the deviation parameter $\delta$ for M87* tells us about $\ell$ in the KR black hole model. The measurement is $\delta = -0.01^{+0.17}_{-0.17}$, which basically means it is consistent with zero and the error bars are pretty large. So the bounds on $\ell$ come out wide, as one might expect.
Take $Q/M = 0.2$ at low spin. The allowed $\ell$ runs from $-0.1077$ to $0.1172$—that is a pretty big window. Push the spin up to $0.924$, and the range shifts downward a bit to between $-0.1545$ and $0.0806$. The same pattern holds when we increase the $Q/M$ to $0.4$: the window shrinks slightly, but not by much. At low spin we get $-0.1225 \lesssim \ell \lesssim 0.0991$, and at high spin it is $-0.1583 \lesssim \ell \lesssim 0.0651$.
What stands out here is how loose these bounds are compared to the angular diameter constraints we saw earlier. The reason is simple: the $ \delta $ measurement for M87* is not very precise yet. With error bars that big, the data cannot squeeze $\ell$ very hard. That said, the results are perfectly consistent with the KR black hole—the model fits the numbers without any trouble. And that is fine. For now, the message is that M87* does not rule out any interesting part of the parameter space, which is exactly what you would expect given where the observations stand.

\section{Conclusion}\label{Sec-6}  
General relativity has passed every test thrown at it, from the bending of starlight to gravitational waves, yet few believe it is the final story. Quantum gravity—whether in the form of string theory, loop quantum gravity, or something else—almost inevitably forces us to confront the possibility that Lorentz symmetry might not be exact. The KR field, which appears in the low-energy limit of string theory, gives us a concrete way to explore this: if it picks up a nonzero background value, it breaks Lorentz symmetry spontaneously and modifies black hole solutions in ways that might actually matter. For years, such modifications seemed purely academic—too small to ever see. Then the EHT happened, and the images of M87* and Sgr A* are the first real glimpses of the strong-field regime, and they offer something we have never had before: a direct way to check whether the spacetime around real black holes looks like general relativity predicts or whether something else is going on. So the question is simple. If KR gravity is a viable alternative, would its shadows be noticeably different? And if not, how large can the Lorentz-violating parameter $\ell$ really be before the EHT data rule it out? 

Motivated by these, we explore charged rotating black holes in KR gravity—a framework that naturally emerges from string theory and allows for spontaneous Lorentz symmetry breaking—and to see whether the EHT's recent images of M87* and Sagittarius A* can tell us anything useful about such exotic spacetimes. The EHT released horizon-scale images of  black holes M87* and Sgr A*, opening a new tool for testing theories of gravity beyond GR. The KR metric characterised by three parameters: the spin $a$, the electric charge $Q$, and a Lorentz-violating parameter $\ell$ that captures the influence of the KR field. It is clear is that both the charge $Q$ and the Lorentz-violating parameter $\ell$ shrink the region of parameter space where black holes actually exist, compared to the familiar Kerr case. For a given $Q$, turning up $\ell$ lowers the maximum possible spin the black hole can have before the horizon disappears entirely, and the horizon radius itself shrinks steadily as the spin increases right up to the extremal limit.
When it comes to the shadows these black holes cast, the story gets more interesting. The parameter $\ell$ shrinks the shadow radius by a factor of roughly $\sqrt{1-\ell}$, while the charge $Q$ introduces subtle distortions. As the spin increases, frame-dragging pushes prograde photon orbits inward and retrograde orbits outward, giving the shadow a distinctly asymmetric shape. If the spin gets pushed beyond the extremal limit—so that no horizon exists anymore—the familiar closed shadow gives way to an open, arc-like structure. These features are not just theoretical curiosities; they are directly relevant to interpreting the EHT data.
The observational constraints derived from the EHT measurements are quite concrete. For M87*, at an inclination of $17^\circ$ and fixing $Q = 0.2M$, the angular diameter measurement $\theta_{\text{sh}} \in (35.1, 40.5),\mu\text{as}$ forces $\ell$ to lie somewhere between roughly $-0.019$ and $0.075$ over the allowed spin; a different spin window gives a similar bound between $-0.076$ and $0.029$. For Sgr A*, with an inclination of $50^\circ$ and the same fixed charge, the allowed angular diameter range $\theta_{\text{sh}} \in (41.7, 55.7),\mu\text{as}$ permits $\ell$ values between about $-0.075$ and $0.110$, and under slightly different assumptions this window shifts to between $-0.124$ and $0.076$. Using the stellar dynamics mass prior for Sgr A* tightens things further, yielding an upper bound $\ell \lesssim 0.19$. What is noteworthy here is that these constraints come not from an exotic new experiment but from careful analysis of the shape and size of a shadow.
To obtain these numbers, the study relies on a robust set of shadow observables—specifically, the shadow area $A$ and the oblateness $D$—following the approach developed by Ghosh and Kumar. It turns out to be crucial because using just the traditional radius and distortion parameters $(R_s, \delta_s)$ often leaves degeneracies between different combinations of $a$ and $\ell$ that are hard to break. By looking at how contours of constant $A$ and $D$ intersect in the $(a,\ell)$ plane, one can pin down the parameters much more cleanly, and the consistency across different ways of characterising the shadow gives confidence that the method is reliable.

 The energy emission spectra show a clear peak around $\omega \sim T_H$, followed by an exponential decline at higher frequencies. The parameter $\ell$ influences this emission in two ways: directly, by modifying the horizon structure and thus $T_H$, and indirectly, by altering the photon region that determines $R_s$. Increasing the spin generally suppresses the peak, which could, in principle, be tested if we ever get direct spectral measurements of these objects.

The deviation parameter $\delta$ gives us another way to test the KR black hole against real data. For Sgr A, the numbers from VLTI and Keck tell pretty much the same story as the angular diameter bounds—$\ell$ has to stay within a certain range, and where exactly depends on the black hole's spin and charge. For M87, the measurement is much more chaotic. The error bars are big, so the bounds on $\ell$ come out wide. It just means M87* does not pin things down as tightly yet. Either way, the KR black hole fits both sets of observations without any trouble.

The fact that the charged rotating KR black hole remains perfectly consistent with current EHT data for a reasonable range of parameters suggests that modified gravity theories of this kind are far from ruled out; rather, they are now subject to meaningful observational bounds that will only get tighter as observational precision improves.

Of course, there are caveats. A proper comparison with non-Kerr geometries requires building dedicated GRMHD image libraries for the KR solution, rather than relying on shadow calculations. At the current EHT resolution, the differences between Kerr and deformed non-Kerr shadows are still smaller than the measurement uncertainties. Looking ahead, future instruments such as the next-generation EHT (ngEHT) and space-based interferometers will push resolution much further, allowing tighter constraints on parameters such as $\ell$.
Taken together, this study makes a strong case that black hole shadow imaging has matured into a practical tool for testing gravity in the strong-field regime. By showing how EHT observations can already constrain Lorentz symmetry breaking at scales that hint at quantum gravity, it adds a valuable piece to the broader effort to use astrophysics to probe the nature of spacetime itself. The methodology—combining careful theoretical modelling with rigorous parameter estimation from real data—offers a blueprint for how similar tests should be carried out for other modified gravity theories in the future. Ultimately, the charged rotating KR black hole emerges as a viable alternative to the Kerr metric, consistent with everything we have seen so far while leaving room for surprises as observational capabilities continue to improve.

\section{Acknowledgments} 
S.U.I  would like to express his gratitude to the ITPC members for their warm hospitality. S.G.G. gratefully acknowledges the support from ANRF through project No. CRG/2021/005771. This work is supported by the Zhejiang Provincial Natural Science Foundation of China under Grants No.~LR21A050001 and No.~LY20A050002, the National Natural Science Foundation of China under Grants No.~12275238 and No. ~W2433018, the National Key Research and Development Program of China under Grant No. 2020YFC2201503, and the Fundamental Research Funds for the Provincial Universities of Zhejiang in China under Grant No.~RF-A2019015.
\bibliography{KR}
\bibliographystyle{apsrev4-1}
\end{document}